\documentclass[12pt,a4paper]{article}

\usepackage{algorithm}
\usepackage{algpseudocode}
\usepackage{geometry}
\usepackage[utf8]{inputenc}
\usepackage[T1]{fontenc}
\RequirePackage{amsmath,amssymb}
\RequirePackage{mathrsfs}
\RequirePackage{amsfonts,bm}
\RequirePackage{dsfont}
\RequirePackage{braket}
\RequirePackage{tabularx}
\usepackage{multirow}
\RequirePackage{nicefrac}
\RequirePackage{graphicx}
\RequirePackage{booktabs}
\RequirePackage{colortbl}
\RequirePackage{xcolor}
\RequirePackage[symbol]{footmisc}
\usepackage{mathtools}
\usepackage{comment}
\usepackage{blkarray}
\usepackage{stackengine}
\usepackage{subcaption}
\usepackage{float}
\usepackage{rotating}
\usepackage{hhline}
\usepackage{tikz}
\usepackage{physics}
\usepackage{multirow}
\usepackage{multicol}
\usepackage[section]{placeins}
\usepackage[natbib=true,maxbibnames=99,giveninits=true,sorting=none]{biblatex}
\usepackage{xurl}
\newenvironment{aline}
    {\begin{equation}
    \begin{aligned}
    }
    { 
    \end{aligned}
    \end{equation}
    \ignorespacesafterend
    }

\def\l{\label}
\def\beqn{\begin{eqnarray}}
\def\eeqn{\end{eqnarray}}

\usepackage{empheq}
\usepackage[most]{tcolorbox}
\usepackage{blkarray}
\newtcbox{\mymath}[1][]{%
    nobeforeafter, math upper, tcbox raise base,
    enhanced, colframe=blue!30!black,
    colback=blue!30, boxrule=1pt,
    #1}

\newcommand{\CC}[2]{C{#1\atopwithdelims[]#2}}
\newcommand{\Z}[2]{Z{#1\atopwithdelims[]#2}}

\newcommand{\ba}{\begin{eqnarray}}
\newcommand{\ea}{\end{eqnarray}}

\newcommand{\vth}{\vartheta} 
\newcommand{\vthb}{\bar{\vartheta}} 
\newcommand{\sqb}[2]{\hspace{-0.02cm}\left[\begin{matrix}#1\\#2\end{matrix} \right]} 
\newcommand{\smb}[2]{\hspace{-0.02cm}\left[\begin{smallmatrix}#1\\#2\end{smallmatrix} \right]}
\newcommand{\smbar}[4]{\hspace{-0.02cm}\left[\begin{smallmatrix}#1\\#2\end{smallmatrix} \middle| \begin{smallmatrix}#3\\#4\end{smallmatrix}  \right]}

\newcommand{\smc}[2]{\hspace{-0.02cm}\left(\begin{smallmatrix}#1\\#2\end{smallmatrix} \right)}

\newcommand{\one}{\mathds{1}} 
\newcommand{\Sv}{\bm{S}} 
\newcommand\e[1]{\bm{e_#1}}
\newcommand\bv[1]{\bm{b_#1}}
\newcommand\z[1]{\bm{z_#1}}
\newcommand\x{\bm{x}}

\numberwithin{equation}{section}

\DeclareMathSymbol{\mg}{\mathrel}{symbols}{"1D}

\renewcommand{\Im}{\text{Im}\ }

\newcommand{\der}{\partial}

\newcommand{\beq}{\begin{equation}}
\newcommand{\eeq}{\end{equation}}
\newcommand{\barr}{\begin{array}}
\newcommand{\earr}{\end{array}}

\newcounter{oldcounter}

\newcommand{\bw}{{\bar w}}

\newcommand{\byy}{{\bar y}}
\newcommand{\bz}{{\bar z}}

\newcommand{\bJ}{{\overline J}}

\newcommand{\bgf}{{\bar\phi}}

\newcommand{\bgps}{{\bar\psi}}
\newcommand{\bget}{{\bar\eta}}

\newcommand{\Bgb}{{\boldsymbol \beta}}

\newcommand{\Intr}{\mathbb{Z}}

\newcommand{\ztwo}{\Intr_2\!\times\!\Intr_2}

\begin{document}
\begin{titlepage}
\samepage{
\setcounter{page}{1}
\rightline{CERN-TH-2023-114}
\rightline{LTH-1342}
\rightline{June 2023}

\vfill
\begin{center}
  {\Large \bf{
     $D$--Term Uplifts \\ \medskip 
     in Non--Supersymmetric Heterotic String Models }}

\vspace{1cm}
\vfill

{\large Alonzo R. Diaz Avalos $^{1}$\footnote{E-mail address: a.diaz-avalos@liverpool.ac.uk}, \\ \medskip
Alon E. Faraggi$^{1,2}$\footnote{E-mail address: alon.faraggi@liverpool.ac.uk}, Viktor G. Matyas$^{1}$\footnote{E-mail address: viktor.matyas@liverpool.ac.uk} and 
 Benjamin Percival$^{1}$\footnote{E-mail address: b.percival@liverpool.ac.uk}}

\vspace{1cm}

{\it $^{1}$ Dept.\ of Mathematical Sciences, University of Liverpool, Liverpool
L69 7ZL, UK\\}
\vspace{.08in}

{\it $^{2}$ CERN, Theoretical Physics Department, CH--1211 Geneva 23, Switzerland\\}

\vspace{.025in}
\end{center}

\vfill
\begin{abstract}
\noindent
Recently, we proposed that the one--loop tadpole diagram in perturbative 
\sloppy non--supersymmetric heterotic string vacua that contain an anomalous 
$U(1)$ symmetry, leads to an analog of the Fayet--Iliopoulos 
$D$--term in $\mathcal{N}=1$ supersymmetric models, and may uplift the vacuum 
energy from negative to positive value. 
In this paper, we extend this 
analysis to new types of vacua, including those with Stringy Scherk--Schwarz (SSS)
spontaneous supersymmetry breaking versus those with explicit breaking. We 
develop a criteria that facilitates the extraction of vacua with
Scherk--Schwarz breaking. We develop systematic tools to analyse 
the T--duality property of some of the vacua and demonstrate them 
in several examples. The extraction of the anomalous $U(1)$ $D$--terms is
obtained in two ways. The first utilises the calculation of the 
$U(1)$--charges from the partition function, whereas the second 
utilises the free fermionic classification methodology to classify 
large spaces of vacua and analyse the properties of the massless 
spectrum. The systematic classification method also ensures that 
the models are free from physical tachyons. We provide a systematic
tool to relate the free fermionic basis vectors and one--loop 
Generalised GSO phases that define the string models, to the 
one--loop partition function in the orbifold representation. We argue that
a $D$--term uplift, while rare, is possible for both the SSS 
class of models, as well as in those with explicit breaking.
We discuss the steps needed to further develop the arguments presented 
here. 

\end{abstract}

\smallskip}

\end{titlepage}

\section{Introduction}\label{intro}

The Standard Model (SM) of particle physics provides an effective parametrisation for
all observational sub--atomic data to date. It is even possible that it remains viable
up to the Grand Unified Theory (GUT) scale, or Planck scale, where
gravitational effects become prominent. In this
case, gaining insight into the fundamental origin of the SM parameters can only be gleaned by fusing it with gravity.
String theory provides the most advanced contemporary framework
to pursue the synthesis of gravity with the sub--atomic gauge
interactions. For that purpose, 
Standard--like Models were constructed in the free fermionic
formulation of the heterotic string and provide a 
laboratory to study how the Standard Model parameters may arise in a
theory of quantum gravity 
\cite{fny,slm1, slm2, cfn, Lebedev:2006kn,Blaszczyk:2010db, slmclass}. 
The free fermionic
heterotic string models are $\ztwo$ orbifolds of six dimensional 
tori at enhanced symmetry points in the moduli space 
\cite{z2z21,KK,z2z23, z2z24,z2z25}. 

While the majority of phenomenological string models constructed to date
possess $\mathcal{N}=1$ spacetime supersymmetry (SUSY), absence of SUSY at
observable energy scales mandates that it has to be broken.
Spacetime SUSY in string models can be broken by non--perturbative
effects in the effective field theory limit of the string vacua,
or it may be broken directly at the string scale. In the string constructions,
we may distinguish between explicit and spontaneous breaking, where, in the former, the
remaining gravitino is projected from the spectrum, whereas spontaneous breaking can 
arise through the Scherk--Schwarz
mechanism \cite{SS2, SS1, SS3, CDC2, CDC1}, 
in which case the gravitino
mass is proportional to the inverse of an internal radius of the
six dimensional compactified torus.

It is clear that addressing many of the questions in string
phenomenology mandates the breaking of SUSY. In
particular when it comes to the cosmological evolution
and string dynamics near the Planck scale. The non--SUSY
string vacua typically contain physical tachyons in their spectra
that indicate that they are unstable. However, also those
configurations that are free of physical tachyons, in general have
non--vanishing tadpoles and vacuum energy that in general lead to
instability. 

One of the recurring features in supersymmetric string derived models
is the existence of an anomalous $U(1)$ ($U(1)_A$) symmetry. The $U(1)_A$ is
cancelled by an analogue of the Green--Schwarz mechanism, which generates
a Fayet--Iliopoulos (FI) $D$--term that breaks SUSY near the Planck
scale \cite{DSW, ads}. The non--vanishing $D$--term gives rise to a non--vanishing
vacuum energy at two--loop \cite{AS}. SUSY
can be restored by assigning non--vanishing Vacuum Expectation
Values (VEV) to some Standard Model singlet fields along $F$-- and
$D$--flat directions. The anomalous $U(1)$ symmetry in string construction
plays a pivotal role in many of the phenomenological studies of string
compactifications \cite{ckm, cfua1, Faraggi:1997be, Faraggi:1998bi}. 

An anomalous $U(1)$ symmetry is a recurring feature also in
non--SUSY string vacua. The same diagram in string
perturbation theory that generates the FI
term in the SUSY configurations is also
present in the non--SUSY configurations, {\it i.e.}
both in those with explicit SUSY breaking,
as well as those in which it is broken by the SSS
mechanism. Similarly, the two--loop diagram contributing to the vacuum
energy is also present in the case of non--SUSY vacua,
either with explicit or SSS SUSY breaking. Thus,
it imperative to take into account this contribution to the vacuum energy
also in these cases.

This contribution to the vacuum energy is particularly pertinent
to the question of the existence of a de--Sitter vacuum in string
theory. Astrophysical and cosmological data indicate that the universe
is accelerating. The existence of a positive vacuum energy is one
of the possible explanations. However, the existence of string
vacua with positive vacuum energy and stable moduli is currently under intense scrutiny and doubt. For instance, for the non--SUSY heterotic constructions examined through effective field theory methods in ref. \cite{Sethi22} only AdS vacua are found to be possible. However, through exact worldsheet evaluation of the one--loop potentials of non--SUSY heterotic string orbifolds, Florakis and Rizos demonstrated
the existence of string vacua with positive vacuum energy \cite{fr1}. However, many open issues remain in the study of string vacua without SUSY and with respect to (related) issues around moduli stabilisation. Additionally, it remains important for this analysis to be extended to models in which more phenomenological
criteria are satisfied. 

One direction in which progress can be made to the evaluation of vacuum energy for non--SUSY heterotic string vacua is through exploring the contribution
of the would--be FI $D$--term. Since this contribution is positive definite it may uplift an \textit{a priori} negative
vacuum energy to a positive one, an idea discussed in \cite{BKQ}. Recently, we demonstrated this
possibility in a particular string vacuum \cite{AFMP}. The analysis utilises
the free fermionic classification methodology to extract
tachyon free non--SUSY string vacua. It then calculates
the traces of the $U(1)$ gauge symmetries and extracts the
tachyon free vacua with a $U(1)_A$. The  
one--loop vacuum amplitude was analysed in comparison to the
$U(1)_A$ would--be $D$--term contribution. Following
refs \cite{fr1,fr2,fr3}, a numerical analysis of the potential
and its dependence on the moduli in
specific string models was performed in the neighbourhood
of the a local minimum. It was then found that a $D$--term
uplift to a positive value may indeed be possible in a model
with explicit supersymmetry breaking. 

In this paper we extend the analysis of \cite{AFMP}.
We develop a criteria to distinguish the models with 
SSS supersymmetry breaking that allow for
the vacuum energy to be exponentially suppressed provided that the number
of massless bosons and fermions is equal. 
We then proceed
to analyse the vacuum energy and potential of a range of models with both explicit and SSS SUSY breaking.
We demonstrate that a $D$--term uplift may indeed be possible in
models with SSS breaking as well as in models with
explicit breaking. We further provide some statistical measure
for the frequency of tachyon free models with SSS breaking,
and provide further examples of cases where a minimum of potential is
not obtained for finite value of the moduli.

Our paper is organised as follows: in section \ref{FFModelBuilding} we review
some general aspects of the free fermion construction that are particularly 
relevant for the analysis in this paper, and refer to the literature for more
details. In section \ref{AnomU1} we review the calculation of the Fayet--Iliopoulos 
term in $\mathcal{N}=1$ supersymmetric string vacua and discuss its adaptation to the 
$\mathcal{N}=0$ case. In section \ref{PFSection} we discuss the analysis of the 
one--loop partition function and potential, as well as the derivation 
of the anomalous $U(1)$ from the partition function that serves as a
counter--check on its derivation from the massless spectrum. 
In section \ref{section:SUSYbreak} we elaborate on explicit SUSY 
breaking versus spontaneous SUSY breaking by the SSS--mechanism. 
We derive conditions that facilitate the extraction of the string 
vacua that utilise the SSS--mechanism and provide examples
that demonstrate their utilisation in Appendix \ref{appendix:SS-T-dualityConditions}. 
Similarly, in section \ref{section:SUSYbreak} we identify the conditions 
on the worldsheet phases that exhibit the T--duality property of the string 
vacua and supplement these with examples in Appendix \ref{appendix:SS-T-dualityConditions}.
In section \ref{section:tachs} we discuss the conditions for the extraction of 
tachyon free configurations in the space of vacua, and section \ref{section:chiralsecs}
elaborates on the analysis of the chiral sectors and extraction of the anomalous 
$U(1)_A$ symmetry from the massless spectrum. Section \ref{results} presents our
results that include examples of uplift models with SSS supersymmetry breaking 
as well as explicit breaking. Section \ref{conclusions} contains our
conclusions and discussion on further steps that can be taken to improve the 
rigour of the analysis presented in this paper as well as its predictability. 
In Appendix \ref{appendix:translation} we discuss in detail how to 
relate a free fermion model which is specified in term of the set of 
boundary condition basis vectors and one--loop GGSO phases, to
the one--loop partition function in a bosonic representation. The art in this
regard is in the translation of the GGSO projection coefficients to the 
modular invariant phase that appears in the one--loop partition function. 
This tool therefore facilitates the writing of the partition function, which 
provides access to the entire string spectrum, for any string model, which is 
specified in terms of the boundary condition basis vectors and one--loop phases.


\section{Model building in the free fermionic formulation}\label{FFModelBuilding}
In the free fermionic formulation of \cite{ABK1,ABK2,KLT}, all degrees of freedom are realised as free fermions propagating on the string worldsheet. For the heterotic string in four dimensions, we consider holomorphic 
fields that realise a supersymmetric $c=10$ conformal algebra and antiholomorphic 
fields that realise a non--supersymmetric conformal algebra. Along with the spacetime bosons $X^\mu(z,\bar{z})$ we denote the fermionic fields  as
\begin{align}
\begin{split}
\text{Holomorphic}: \ \ \ &\psi^{\mu=1,2},\chi^{1,...,6},y^{1,...,6},w^{1,..,6}\ (z)\\
\text{Antiholomorphic}: \ \ \ &\bar{y}^{1,...,6},\bar{w}^{1,..,6},\bar{\psi}^{1,...,5},\bar{\eta}^{1,2,3},\bar{\phi}^{1,...,8}\ (\bar{z}),
\end{split}
\end{align}
where: 
\begin{itemize}
    \item $\psi^{\mu}$ is the superpartner of the bosonic spacetime field $X^\mu(z)$ in the lightcone gauge.
    \item $\chi^{1,...,6}$ are the superpartners of the six compact directions of the bosonic coordinate fields.
    \item $\{y^i,w^i \ | \ \bar{y}^i,\bar{w}^i \}$ realise the conformal field theory associated to the six dimensional compact geometry.
    \item Currents associated to $\{\bar{\psi}^{1,...,5},\bar{\eta}^{1,2,3}\}$ can realise a gauge $c=8$ conformal algebra containing an $SO(10)$ GUT from the $\bar{\psi}^{1,...,5}$ that may contain Standard Model like gauge fields. 
    \item Currents associated to $\{\bar{\phi}^{1,...,8}\}$ can be associated to a gauge $c=8$ conformal algebra relating to the hidden sector of the theory.
\end{itemize} 
Consistent model building requires that a $N=1$ superconformal algebra on the string worldsheet is realised among the holomorphic degrees of freedom. This can be achieved via the worldsheet supercurrent
\beq \label{scurrent}
T_F(z)=i\psi^\mu \partial X^\mu (z)+i\sum^6_{I=1}\chi^I y^I w^I,
\eeq 
which has conformal weight $(3/2,0)$. This results in a local enhanced symmetry group $SU(2)^6$, the adjoint representation of which is given by the six $SU(2)$--triplets from $\{\chi^I,y^I,w^I\}$.

Models in the free fermionic formalism are then defined by considering a one--loop torus and defining a set of $N$ boundary condition basis vectors, $\bm{v_i}\in \mathcal{B}$, specifying how each free fermion, $f$, propagates around the two non--contractible loops of the torus. An element $\Bgb$ of the space $\Xi=\text{span}\{\mathcal{B}\}$ can then be written as
\begin{align}
\begin{split}
\Bgb=\{&\beta(\psi^\mu),\beta(\chi^{12}),\beta(\chi^{34}),\beta(\chi^{56}),\beta(y^1),...\beta(w^6)\ | \ \\ 
&\beta(\byy^1),...,\beta(\bw^6); \ \beta(\bgps^{1}),...,\beta(\bgps^{5}),\beta(\bget^{1}),\beta(\bget^{2}),\beta(\bget^{3}),\beta(\bgf^{1}),...,\beta(\bgf^{8})\},
\end{split}
\end{align}
such that $\beta(f)\in (-1,1]$ and Ramond (R) boundary conditions correspond to $\beta(f)=1$, while Neveu--Schwarz (NS) is given by $\beta(f)=0$.

The partition function in the fermionic formulation can be written as
\beq \label{FFPF}
Z=Z_B\sum_{\bm{\alpha},\bm{\beta}\in\Xi} \CC{\bm{\alpha}}{\bm{\beta}} \Z{\bm{\alpha}}{\bm{\beta}},
\eeq 
where $Z_B=1/\eta^2\bget^2$ is the bosonic partition function and $\CC{\bm{\alpha}}{\bm{\beta}}$ are Generalised GSO (GGSO) phases which respect modular invariance. The $Z  {\bm{\alpha} \atopwithdelims [] \bm{\beta}}$ represent the worldsheet fermions and are thus products of Jacobi theta functions. The partition function for the models we explore in this work is discussed in section \ref{PFSection}.

Aside from the partition function we can also view the spectrum through the modular invariant Hilbert space, $\mathcal{H}$, of states, $\ket{S_{\Bgb}}$. This is constructed through implementing the one--loop GGSO projections on each sector according to:
\begin{equation}
    \mathcal{H}=\bigoplus_{\Bgb\in \Xi}\prod^{N}_{i=1}
    \left\{ e^{i\pi \bm{v_i}\cdot F_{\Bgb}}\ket{S_{\Bgb}}=\delta_{\Bgb}
    \CC{\Bgb}{\bm{v_i}}^*
    \ket{S_{\Bgb}}\right\}\mathcal{H}_{\Bgb},
\end{equation}
where $F_{\Bgb}$ is the fermion number operator and $\delta_{\Bgb}$ is the spin--statistics index.

The sectors, $\Bgb$, in the model can be characterised according to their holomorphic (H) and
antiholomorphic (A) moving vacuum separately
\begin{align}\label{massform}
\begin{split}
M_H^2&=-\frac{1}{2}+\frac{\Bgb_H \cdot\Bgb_H}{8}+N_H\\
M_A^2 &=-1+\frac{\Bgb_A \cdot \Bgb_A}{8}+N_A,
\end{split}
\end{align}
where $N_H$ and $N_A$ are sums over left and right moving oscillator frequencies,
respectively
\begin{align}
    N_H&=\sum_{\lambda}\nu_\lambda+\sum_{\lambda^*}\nu_{\lambda^*} \\
    N_A&=\sum_{\bar{\lambda}}\nu_{\bar{\lambda}}+\sum_{\bar{\lambda}^*}\nu_{\bar{\lambda}^* },
\end{align}
where $\lambda$ is a holomophic oscillator and $\bar{\lambda}$ is an antiholomorphic oscillator and the frequency is defined through the boundary condition in the sector $\Bgb$
\beq \label{freq}
\nu_\lambda = \frac{1+\beta(\lambda)}{2}, \ \ \ \nu_{\lambda^*} = \frac{1-\beta(\lambda)}{2}.
\eeq 
Physical states must satisfy the
Virasoro matching condition, $M_H^2=M_A^2$, such that massless states are those with $M_H^2=M_A^2=0$ and on--shell tachyons arise for sectors with $M_H^2=M_A^2<0$. 

\subsection{Symmetric $\ztwo$ $SO(10)$ models}
For this work we explore the one--loop cosmological constant and $U(1)_A$ tadpole calculations for models defined through the basis set
\begin{align}\label{SO10basis}
{\one}&=\{\psi^\mu,\
\chi^{1,\dots,6},y^{1,\dots,6}, \omega^{1,\dots,6}\ |   ~~~\overline{y}^{1,\dots,6},\overline{\omega}^{1,\dots,6},
\overline{\eta}^{1,2,3},
\overline{\psi}^{1,\dots,5},\overline{\phi}^{1,\dots,8}\},\nonumber\\
\Sv&=\{{\psi^\mu},\chi^{1,\dots,6}\},\nonumber\\
\e{i}&=\{y^{i},w^{i}\; | \; \overline{y}^i,\overline{w}^i\},
\
i=1,\dots,6,\nonumber\\
\bv{1}&=\{\chi^{34},\chi^{56},y^{34},y^{56}\; | \; \overline{y}^{34},
\overline{y}^{56},\overline{\psi}^{1,\dots,5},\overline{\eta}^1\},\\
\bv{2}&=\{\chi^{12},\chi^{56},y^{12},y^{56}\; | \; \overline{y}^{12},
\overline{y}^{56},\overline{\psi}^{1,\dots,5},\overline{\eta}^2\},\nonumber\\
\z{1}&=\{\overline{\phi}^{1,\dots,4}\},\nonumber\\
\z{2}&=\{\overline{\phi}^{5,\dots,8}\}.
\nonumber
\end{align}
Such a basis can be associated with symmetric $\mathbb{Z}_2\times \mathbb{Z}_2$ orbifolds \cite{z2z21,KK,z2z23,z2z24,z2z25} extensively classified in previous works \cite{slmclass,class1,fknr,fkr1,acfkr2,su4,su62,frs,lrsclass}. The NS sector of the models associated to this basis produce spacetime vector bosons generating the gauge group
\beq 
SO(10)\times U(1)_1\times U(1)_2\times U(1)_3\times SO(8)^2,
\label{eqn:GaugeGroup}
\eeq 
where we note that $U(1)_{1,2,3}$ are generated by the antiholomorphic currents $\bar{\eta}^k\bar{\eta}^{k*}$. With respect to the basis (\ref{SO10basis}) it is useful to identify the important linear combination
\beq 
\x = \one + \Sv + \sum_{i=1}^6 \e{i} +\z{1}+\z{2}=\{\bar{\eta}^{1,2,3},\bar{\psi}^{1,...,5}\}
\eeq 
and $\bv{3}=\bv{1}+\bv{2}+\x=\{\chi^{12},\chi^{34},y^{12},y^{34}\; | \; \overline{y}^{12},
\overline{y}^{34},\overline{\psi}^{1,\dots,5},\overline{\eta}^3\}$, which spans the third twisted plane of the $\mathbb{Z}_2\times \mathbb{Z}_2$ orbifold and facilitates the analysis of the observable spinorial and vectorial representations as first developed in \cite{class1,fknr}. 

Models may then be defined through the choice of GGSO phases $\CC{\bm{v_i}}{\bm{v_j}}$. There are 66 free phases for this basis, with all others specified by modular invariance. The full space of models is thus of size $2^{66}\sim 10^{19.9}$. The $\mathcal{N}=1$ supersymmetric subset of which is defined by those satisfying 
\beq 
\CC{\bm{S}}{\e{i}}=\CC{\bm{S}}{\bm{z_1}}=\CC{\bm{S}}{\bm{z_2}}=-1,
\label{eqn:ProjectorGravitino}
\eeq
in order to preserve one gravitino. Furthermore, we note that the phases $\CC{\mathds{1}}{\bm{S}}$ and $\CC{\bm{S}}{\bm{b_k}}$, $k=1,2,3$, determine the chirality of the degenerate Ramond vacuum $\ket{\bm{S}}$ and the gravitino is retained so long as
\beq \label{eqn:SUSYChiral}
\CC{\mathds{1}}{\bm{S}}=\CC{\bm{S}}{\bm{b_1}}\CC{\bm{S}}{\bm{b_2}}\CC{\bm{S}}{\bm{b_3}},
\eeq 
which can, without loss of generality, be fixed to
\beq \label{SUSYChiral2}
\CC{\mathds{1}}{\bm{S}}=\CC{\bm{S}}{\bm{b_1}}=\CC{\bm{S}}{\bm{b_2}}=-1, 
\eeq 
for a scan of $\mathcal{N}=1$ vacua. 

Since we are interested in non--SUSY vacua in this work we will be considering the complement to this space of $\mathcal{N}=1$ vacua. In previous work on non--SUSY heterotic string vacua from $\mathbb{Z}_2\times \mathbb{Z}_2$ orbifolds \cite{so10tclass,PStclass,asymmclass} tachyon free configurations satisfying various phenomenological requirements and their one--loop cosmological constants are explored. In section \ref{section:tachs} we will detail how we ensure that only those models free from physical tachyons are explored. 

In the next section we show how the $U(1)_A$ gauge transformation manifests in a 4 dimensional theory and how the Green--Schwarz mechanism deals with it. We then compute the FI term through a string theory computation at 1--loop and we conjecture how this term can lift the vacua from an anti--de Sitter to a de Sitter.

\section{Fayet--Iliopoulos $D$--term calculation}\label{AnomU1}
Anomalies arise whenever one, or some, of the classical symmetries are broken by quantum effects. Some global symmetries need to be broken by anomalies in order
to reproduce observable phenomenology, however a breakdown of a local symmetry indicates a symmetry current is no longer conserved and longitudinal, non--physical modes of the gauge fields no longer may decouple from the S--matrix. This can result
in the loss of unitarity and appearance of unphysical divergences.

In a four dimensional heterotic string theory the anomalies come from the one--loop triangle diagram, 
where the external lines can be gauge fields, gravitons or a mixture of each. In the following, for simplicity, we will consider purely $U(1)$ gauge anomalies.  \\
\noindent
We start from the four dimensional effective action in the Einstein frame
\begin{equation}
    S = \frac{1}{2 \kappa^2} \int d^{4} x \sqrt{-G} \left[ - \frac{\kappa^2 e^{-2 \Phi}}{2 g^2} F_{\mu \nu} F^{\mu \nu} - \frac{e^{-4 \Phi}}{12} H_{\mu \nu \rho} H^{\mu \nu \rho} \right]
\end{equation}
with $H$ the field strength of the $B$ field such that
\begin{equation}
    H = dB - \frac{\kappa^2}{g^2} \Omega_3 ^{YM} = dB - \frac{\kappa^2}{g^2} \Tr \left[ A \wedge dA - \frac{2i}{3} A \wedge A \wedge A \right].
\end{equation}
Under a Gauge transformation, the gauge fields transform as $A_\mu \longrightarrow A_\mu + \partial_\mu \Lambda$ and the effective action varies as follows
\begin{equation}
    \delta S_{one-loop} = \frac{1}{4} \frac{1}{96 \pi ^2} \int d^4 x \Tr [Q^3] \epsilon^{\mu \nu \rho \sigma} \Lambda F_{\mu \nu} F_{\rho \sigma}. 
\end{equation}
When the sum of these $U(1)$ charges is not zero this anomalous triangle diagram contribution will be present, with massless particles circulating in the loop. \\  
\noindent
The Green--Schwarz mechanism \cite{Green:1984sg} provides a way to cancel these one loop anomalies through the introduction of an antisymmetric $2$--form coupled at one loop to the $U(1)$ $2$--form field strength in the effective Lagrangian 
\begin{equation}
    -\frac{\zeta}{2}  \epsilon^{\mu \nu \rho \sigma} B_{\mu \nu} F_{\rho \sigma},
\end{equation}
imposing the $B$ field to vary under the $U(1)$ Gauge transformation as $\delta B = \frac{\kappa ^2}{g^2} \Lambda F$, with the condition $4 \Tr[Q^3] = \Tr[Q]$, such that it compensates the anomalous triangle diagram. \\
\noindent
The action can also be written in terms of the axion field, $a$, dual to the antisymmetric $B$ field. The axion field is introduced as a Lagrange multiplier term into the Lagrangian 
\begin{align}
\begin{split}
    \frac{1}{\sqrt{-G}}\mathcal{L} =& - \frac{e^{-2 \Phi}}{4 g^2} F_{\mu \nu} F^{\mu \nu} - \frac{e^{-4 \Phi}}{24 \kappa^2} H_{\mu \nu \rho} H^{\mu \nu \rho} - \frac{\zeta}{2} \epsilon^{\mu \nu \rho \sigma} B_{\mu \nu} F_{\rho \sigma} \\
    &+ \frac{a}{6} \epsilon^{\mu \nu \rho \sigma} \partial_\mu H_{\nu  \rho \sigma} + \frac{a}{4} \frac{\kappa^2}{g^2} \epsilon^{\mu \nu \rho \sigma} F_{\mu \nu} F_{\rho \sigma} , 
    \label{eq:Lagrangian-axion}
    \end{split}
\end{align}
such that its equations of motion give the definition of the $H$ tensor
\begin{align}
\begin{split}
    &\frac{1}{6} \epsilon^{\mu \nu \rho \sigma} \partial_\mu H_{\nu \rho \sigma} + \frac{\kappa^2}{4 g^2} \epsilon^{\mu \nu \rho \sigma} F_{\mu \nu} F_{\rho \sigma} = 0 \\
    \implies& H = dB - \frac{\kappa^2}{g^2} A \wedge F.
\end{split}
\end{align}
Using this result and integrating out the $H$ tensor in the Lagrangian (\ref{eq:Lagrangian-axion}) gives
\begin{equation}
    \frac{1}{\sqrt{-G}}\mathcal{L} = - \frac{e^{-4 \Phi}}{4 g^2} F_{\mu \nu} F^{\mu \nu} - \kappa^2 e^{4 \Phi} \left( \partial_\mu a + 2 \zeta A_\mu\right)^2 + \frac{a}{4} \frac{\kappa^2}{g^2} \epsilon^{\mu \nu \rho \sigma} F_{\mu \nu} F_{\rho \sigma},
\end{equation}
while the gauge transformation of the axion field can be found through the gauge invariance of $H$ to give
\begin{equation}
    \delta a = - 2 \zeta \Lambda,
\end{equation}
such that when $\zeta \neq 0$ the anomalous gauge field acquires a mass and the gauge symmetry is spontaneously broken. 

In order to compute the FI coefficient $\zeta$ we have to evaluate the 2--point function of the antisymmetric $B$ field and the $U(1)_A$ gauge boson at one loop, where only chiral fermions with odd spin structure give a non--vanishing contribution.

For the ghosts $b,c$ their zero modes are saturated by inserting $\left< b \bar{b} c \bar{c} \right>$ into the path integral. 
For the ghost superpartners $\beta, \gamma$ in order to project out of integration their zero modes one of the vertex operators has to be put in the $0$ picture, the other in the $-1$ and an insertion of the picture changing operator $e^{\phi} T_F$ is needed, where the scalar $\phi$ arises by the bosonization of the $\beta-\gamma$ fields
\begin{equation}
    \beta = e^{-\phi} \partial \xi, \qquad \gamma = e^\phi \eta.
\end{equation}
The amplitude then reads
\begin{equation}
    \int \frac{d \tau_1 d\tau_2}{2 \tau_2} \int d^2 z d^2w \: \left< b(0) \bar{b}(0) c(0) \bar{c}(0) \: e^{\phi} T_F(0) \: V_{A,0} ^{\mu}(z) V_{B,-1} ^{\nu \rho}(w) \right>
\end{equation}
with the denominator needed to fix the translation Killing symmetries of the torus and the residual discrete symmetry $z \longrightarrow -z$.
The vertex operators for the gauge boson and the antisymmetric field in the $-1$ and $0$ pictures are the following
\begin{equation}
    \begin{split}
        V_{A,-1}^{\mu,i} &= g_c \hat{k}^{-1/2} \psi^\mu \bar{J}^i e^{ik \cdot X}\\
        V_{A,0}^{\mu} &= \sqrt{\frac{2}{\alpha'}}g_c \hat{k}^{-1/2} \left( i \partial X^\mu + \frac{\alpha'}{2} k\cdot \psi \psi^\mu \right) \bar{J} e^{ik \cdot X}\\
        V_{B,-1}^{\mu \nu} &= i \sqrt{\frac{2}{\alpha'}}g_c \bar{\partial} X^\mu e^{-\phi} \psi^\nu e^{ip \cdot X}\\
        V_{B,0}^{\mu \nu} &= \frac{2i}{\alpha'} g_c \bar{\partial} X^\mu \left( i \partial X^\nu + \frac{\alpha'}{2} p\cdot \psi \psi^\nu \right) e^{ik \cdot X},
    \end{split}
\end{equation}
with $g_c = \kappa/2 \pi$ and $\hat{k}= 1/2$. These normalisations are chosen such that the string amplitudes match with the field theory calculation and with the vertex operators in the two pictures related via $V_{q+1}(z) = \lim\limits_{w \to z} e^{\phi} T_F (w) V_q (z)$, where the matter supercurrent takes the form
\beq 
T_F = i \sqrt{\frac{2}{\alpha'}} \psi^\mu \partial X_\mu.
\eeq 
We note that we included the $\alpha'$ dependence and that the supercurrent is composed of only the uncompactified fields since vertex operators do not involve internal lattice excitations. \\
\noindent
When $\mathcal{N}\geq 2$ the amplitude vanishes due to the fermion zero modes. However, for $\mathcal{N}= 1$ they can be soaked up by the fermion correlator. To see this, we can take the $\mathcal{O}(p)$ linear approximation of the amplitude
\beq
    - \frac{k_\alpha}{\alpha^{'1/2}}g_c ^2 \int \frac{d \tau_1 d\tau_2}{ \tau_2}  \int d^2 z d^2w \: \langle b \bar{b} c \bar{c} \rangle \: \langle \psi^\sigma  \psi^\alpha \psi^\mu  \psi^\rho \rangle \: \langle \bar{J}^i \rangle \: \langle \partial X^\gamma \bar{\partial}X^\nu \rangle \: \langle e^{\phi} e^{-\phi}\rangle \eta_{\sigma \gamma},
\label{equation:VertexOperatorCalculation}
\eeq
where the fermion correlator can be written in terms of the fermion current correlator 
\begin{equation}
    \langle \psi^i \psi^j \rangle = \langle J^{ij} \rangle = \frac{\epsilon^{ij}}{2 \pi i} \partial_\nu \Tr\left[ (-1)^F e^{2 \pi i \nu J^{ij}} \right]_{\nu = 0}, 
\end{equation}
such that the $\langle \psi^\sigma \psi^\alpha \psi^\mu \psi^\rho \rangle$ term acts as follows
on the four non--compact fermions
\begin{equation}
    \langle \psi^\sigma \psi^\alpha \psi^\mu \psi^\rho \rangle = \epsilon^{\sigma \alpha \mu \rho} \frac{\partial_\nu}{2 \pi i}\frac{\partial_\omega}{2 \pi i} \left( - \frac{1}{2} \frac{\vartheta_1(\nu)}{\eta^2}\frac{\vartheta_1(\omega)}{\eta} \right) = \frac{1}{2} \epsilon^{\sigma \alpha \mu \rho} \eta^4,
\end{equation}
making use of $\partial_\nu \vartheta_1 (\nu) = 2 \pi \eta^3$. The other one--loop correlators are the following
\begin{equation}
    \begin{split}
        \langle b \bar{b} c \bar{c} \rangle &= \eta^2 \bar{\eta}^2 \\
        \langle  X^\gamma(w,\bar{w}) X^\nu (0) \rangle &= \left(- \frac{\alpha'}{2} \ln \vartheta_1 \left( \frac{w}{2 \pi}\right)\bar{\vartheta}_1 \left( \frac{\bar{w}}{2 \pi}\right)  + \frac{\alpha'}{4 \pi \tau_2} \left( \Im z \right)^2 \right) \eta^{\gamma \nu} \\
        \langle e^{\phi} e^{-\phi}\rangle &= \frac{1}{\eta^2},
    \end{split}
\end{equation}
and for the current
\begin{equation}
    \langle \bar{J}^i \rangle = \frac{\partial_{\nu}}{2 \pi i} \Tr \left[ \left( -1 \right)^{F}  q^H \bar{q}^{\bar{H}} e^{2 \pi i  \nu \bar{J}^i}  \right]\Big|_{\nu = 0} = \bar{\eta}^2  \Tr \left[ \left( -1 \right)^{F}  q^H \bar{q}^{\bar{H}- \frac{1}{12}} q_i \right],
\end{equation}
where the derivative acts on the 
the states charged under $U(1)_i$, $i=1,2,3$. In the massless limit only chiral fermions contribute with $\left(H, \bar{H}\right) = \left( 0, \frac{1}{12}\right)$ so we get
\begin{equation}
    \langle \bar{J}^i \rangle = \bar{\eta}^2 \Tr\left[ \left(-1\right)^F q_i \right] = \bar{\eta}^2 \Tr Q_i,
\end{equation}
with $\Tr Q_i= \sum n \, q_{i} \, h$ summed over all states in the spectrum such that $n$, $q_{i}$ and $h$ are the number of massless fermions, their $U(1)_i$ charges and their chirality, respectively.\\
Note that the anomalous $U(1)_A$ charge is given as combination of the three $U(1)_{1,2,3}$ produced by the world--sheet currents $: \bar{\eta}^{i*} \bar{\eta}^i :$, $i=1,2,3$, according to
\beq 
U(1)_A=
\sum_{i=1}^3 \frac{aU_1+bU_2+cU_3}{\sqrt{a^2+b^2+c^2}}
\label{eqn:AnomalousU1}
\eeq 
where $(a,b,c)=\frac{1}{k}(\Tr U_1,\Tr U_2,\Tr U_3)$ and $k=\text{gcd}(\Tr U_1,\Tr U_2,\Tr U_3)$. \\
\noindent
Adding the contribution of the four non--compact bosons from the partition function and their zero modes 
allows us to write the amplitude as
\begin{equation}
    \frac{i g_c^2}{256 \pi^5 \alpha^{'3/2}} k_\alpha \epsilon^{\nu \alpha \mu \rho} \Tr Q_A \int \frac{d \tau_1 d\tau_2}{ \tau_2 ^4}  \int d^2 z d^2w = \frac{i g_c^2}{12 \alpha^{'3/2}} k_\alpha \epsilon^{\nu \alpha \mu \rho} \Tr Q_A,
\end{equation}
using that $\int d^2z = 2 \left( 2 \pi \right)^2 \tau_2$ and $\pi/3$ is the volume of the fundamental domain. 
Reducing the constants and doing further contractions of this $\mathcal{N}=1$ amplitude allows us to finally write the  
FI term as
\begin{equation}
    \zeta = M_s ^2\frac{\Tr Q_A}{192 \pi^2},
    \label{eqn:FI}
\end{equation}
such that, when $\Tr Q_A\neq 0$, there is an additional positive contribution in the effective $D$--term potential
\begin{equation}
    V_D = \frac{1}{2} g_s ^2 \zeta^2
\label{eq:VD}
\end{equation}
that corresponds to a two--loop dilaton tadpole.
This additional term was originally computed both from the low energy energy effective action in ref. \cite{DSW}, 
as well as through explicit two--loop string calculations \cite{AS}.

As stated above, only for $\mathcal{N}=1$ unbroken supersymmetry is the FI term non--vanishing. In the free fermionic models, the breaking $\mathcal{N}=4 \rightarrow \mathcal{N} =1$ is achieved by the introduction of the $\bm{b_1}$ and $\bm{b_2}$ vectors (\ref{SO10basis}) associated to $\mathbb{Z}_2 \times \mathbb{Z}_2$ orbifold twists. The breaking of the last supersymmetry $\mathcal{N}=1 \rightarrow \mathcal{N}=0$ is achieved by setting properly the GGSO phases as delineated in section 2. We will discuss the contribution of the FI $D$--term associated with $\Tr U(1)_A$ in these $\mathcal{N}=0$ models in the following sections. 

Our statement is that once the last supersymmetry is broken, either spontaneously or explicitly, the $D$--term contribution (\ref{eq:VD}) will still be present. 
The same tadpole diagram leading to the FI term in the $\mathcal{N}=1$ supersymmetric case is also present in the $\mathcal{N}=0$ case. The analysis outlined above follows through irrespective of whether the model has $\mathcal{N}=1$ or
$\mathcal{N}=0$ supersymmetry. The supersymmetric case does guarantee a measure of stability whereas the non--supersymmetric case is fraught with further uncertainties. 
For example, non--supersymmetric string vacua contain additional tadpole diagrams 
for the dilaton, which indicate that the string equations of motion are not 
satisfied in Minkowski four dimensional spacetime with constant dilaton, 
and the computational stability from higher loops is not preserved. 
Furthermore, in the following, analysis of the potential is performed 
with respect to a single internal moduli and all other internal moduli 
are set at the free fermionic point but are not fixed. One direction of improvement
on the analysis that we present here is to use the Kiritsis--Kounnas modular 
invariant regularisation scheme \cite {KKregular} that can regulate the infrared divergences. 
We note that these caveats are relevant in general in non--supersymmetric 
string vacua that have been of some old and recent interest in the literature 
\cite{DH, AGMV, KLTClass, nonsusy2, ADM, nonsusy5, aafs, CoCSuppression2, nonsusy9}. 
For our purposes here we note that the contribution of eq. (\ref{eq:VD}),
when non--zero, which has the same mass dimension, $M_s ^4$, as the cosmological constant, will destabilize the vacua adding a positive contribution and lift the minima of the one--loop potential.

For the purpose of getting numerical results we fix the string coupling $\mathcal{O}(g_s)\sim 1$, which corresponds to $\alpha' = g_s / 4\pi \sim 0.1$. This order of magnitude can be justified by reference to work on gauge coupling unification from string model building, for example in ref. \cite{Dienes:1995bx}. 
We observe that the $D$--term contribution goes with $g_s^2$ so smaller values will quickly make an uplift less likely. In section \ref{results}, we will see that we obtain an uplift from AdS to dS only very rarely in our setup and so choosing this order of magnitude for the string coupling, rather than a smaller one, helps provide a proof of concept. 

In the next sections we study some heterotic string models through the analysis of the partition function and its potential behavior and we explicitly show how the FI term is used in order to uplift the minima. 

\section{Partition function and one--loop potential}\label{PFSection}
The generic form of the partition function is given in eq. (\ref{FFPF}) applied to the models defined through the basis (\ref{SO10basis}). Using the techniques developed in Appendix \ref{appendix:translation}, the partition function written in the free fermionic construction can then be written in the following form 
\begin{aline}\label{PF}
    Z=\,&\frac{1}{\eta^{10}\bar{\eta}^{22}} \, \frac{1}{2^{2}} \sum_{\substack{a,k\\b,l}} \;\frac{1}{2^{6}} \,\sum_{\substack{H_i\\G_i}} \;\frac{1}{2^{4}}\sum_{\substack{h_1,h_2,P_i\\g_1,g_2,Q_i}} (-1)^{a+b+P_1 Q_1 + P_2 Q_2+\Phi\smb{a&k&H_i&h_1&h_2&P_i}{b&l&G_i&g_1&g_2&Q_i}}\\[0.2cm]
    &\times \vth\smb{a}{b}_{\psi^\mu} \vth\smb{a+h_1}{b+g_1}_{\chi^{12}} \vth\smb{a+h_2}{b+g_2}_{\chi^{34}} \vth\smb{a-h_1-h_2}{b-g_1-g_2}_{\chi^{56}}\\[0.3cm]
    &\times \Gamma^{(1)}_{2,2}\smbar{H_1&H_2}{G_1&G_2}{h_1}{g_1}(T^{(1)},U^{(1)}) \\[0.2cm]
    &\times \Gamma^{(2)}_{2,2}\smbar{H_3&H_4}{G_3&G_4}{h_2}{g_2}(T^{(2)},U^{(2)}) \\[0.2cm]
    &\times \Gamma^{(3)}_{2,2}\smbar{H_5&H_6}{G_5&G_6}{h_1+h_2}{g_1+g_2}(T^{(3)},U^{(3)}) \\[0.3cm]
     &\times \vthb\smb{k}{l}^5_{\bar{\psi}^{1-5}} \vthb\smb{k+h_1}{l+g_1}_{\bar{\eta}^1} \vthb\smb{k+h_2}{l+g_2}_{\bar{\eta}^2} \vthb\smb{k-h_1-h_2}{l-g_1-g_2}_{\bar{\eta}^3} \vthb\smb{k+P_1}{l+Q_1}^4_{\bar{\phi}^{1-4}} \vthb\smb{k+P_2}{l+Q_2}^4_{\bar{\phi}^{5-8}}.
\end{aline}
For symmetric $\ztwo$ orbifolds the moduli space is generally parametrised by three complex structure and three K\"ahler moduli, one for each torus associated to the $\Gamma_{2,2}$ lattices. The moduli space is then $SO(2,2)/SO(2)\times SO(2)$. 
At the maximally symmetric (free fermionic) point $(T_*, U_*)$, at which bosonic degrees of freedom can be fermionised, the lattices admit a factorised form which can be written entirely in terms of theta functions
\begin{align} \label{GammaFF}
    & \Gamma^{(1)}_{2,2}\smbar{H_1&H_2}{G_1&G_2}{h_1}{g_1}(T^{(1)}_*,U^{(1)}_*)  = \left| \vth\smb{H_1}{G_1} \vth\smb{H_1+h_1}{G_1+g_1} \vth\smb{H_2}{G_2} \vth\smb{H_2+h_1}{G_2+g_1} \right| \nonumber \\[0.2cm] 
    &\Gamma^{(2)}_{2,2}\smbar{H_3&H_4}{G_3&G_4}{h_2}{g_2}(T^{(2)}_*,U^{(2)}_*) = \left| \vth\smb{H_3}{G_3} \vth\smb{H_3+h_2}{G_3+g_2} \vth\smb{H_4}{G_4} \vth\smb{H_4+h_2}{G_4+g_2} \right|\\[0.2cm]
    & \Gamma^{(3)}_{2,2}\smbar{H_5&H_6}{G_5&G_6}{h_1+h_2}{g_1+g_2}(T^{(3)}_*,U^{(3)}_*) = \left| \vth\smb{H_5}{G_5} \vth\smb{H_5-h_1-h_2}{G_5-h_1-h_2} \vth\smb{H_6}{G_6} \vth\smb{H_6-h_1-h_2}{G_6-h_1-h_2} \right|. \nonumber
\end{align}
We furthermore note that the modular invariant phase $\Phi\smb{a&k&H_i&h_1&h_2&P_i}{b&l&G_i&g_1&g_2&Q_i}$ in (\ref{PF}) implements the various GGSO projections. A choice of phase is equivalent to a choice of GGSO matrix and hence there is a unique one--to--one map between them. The factor of $a+b$ ensures correct spin statistics, while the explicit inclusion of the extra phase $P_1 Q_1 + P_2 Q_2$ means that $\Phi=0$ is a valid modular invariant choice.

The summation indices used to write the fermionic partition function \eqref{PF} correspond to various features of the model. The indices $a,b$ correspond to the spin structures of the spacetime fermions $\psi^\mu$, while $k,l$ are associated to the 16 right--moving complex fermions giving the gauge degrees of freedom of the heterotic string. The non--freely acting $\mathbb{Z}_2\times \mathbb{Z}_2$ orbifold twists are associated to the parameters $h_1,g_1$ and $h_2,g_2$. One of the key features of models defined by the basis \eqref{SO10basis} is the inclusion of the basis vectors $\e{i}$ which generate freely acting orbifold shifts in the internal dimensions of the compact torus. In the partition function, these are realised by the indices $H_i,G_i$ parametrising each of the six independent shifts. The additional twists $P_i,Q_i$ correspond to the basis vectors $\z{1}$ and $\z{2}$ acting on the hidden sector of our model.

The moduli--dependent form of the twisted/shifted lattice requires closer attention. We know that all dependence on the geometric moduli is contained in the untwisted sector of the model and hence
\begin{equation}\label{Z22Twisted}
    \Gamma_{2,2}\smbar{H_1&H_2}{G_1&G_2}{h}{g}(T,U)\Big|_{h,g\neq0} = \Gamma_{2,2}\smbar{H_1&H_2}{G_1&G_2}{h}{g}(T_*,U_*).
\end{equation}
This means that for nonzero twists the lattice is precisely given by its factorised form in \eqref{GammaFF}. In the case of the untwisted sector, the shifted lattice can be written in a Poisson resummed Hamiltonian form as 
\begin{equation}\label{Z22TU}
   \Gamma_{2,2}\smbar{H_1&H_2}{G_1&G_2}{0}{0}(T,U) = \sum_{m_i,n_i\in\mathbb{Z}} q^{\frac{1}{2} |\mathcal{P}_L(T,U)|^2} \bar{q}^{\frac{1}{2} |\mathcal{P}_R(T,U)|^2} e^{i\pi\sum_i((m_i+n_i+H_i)G_i},
\end{equation}
where the left and right--moving momenta are 
\begin{aline}\label{PLPR}
    \mathcal{P}_L =& \frac{1}{\sqrt{2 T_2 U_2}}\left[ \frac{U}{2}(m_1+n_1) - \frac{1}{2}(m_2+n_2) + T(m_1-n_1+H_1) + TU(m_2-n_2+H_2) \right] \\
    \mathcal{P}_R =& \frac{1}{\sqrt{2 T_2 U_2}}\left[ \frac{U}{2}(m_1+n_1) - \frac{1}{2}(m_2+n_2) + \bar{T}(m_1-n_1+H_1) - TU(m_2-n_2+H_2) \right].
\end{aline}
Written in this form, it is easy to extract the $q$--expansion of the partition function at any given point in the moduli space which is crucial for calculating the one--loop potential. It can be shown that the twisted/shifted lattice sums \eqref{Z22Twisted} and \eqref{Z22TU} evaluated at the special point $(T_*,U_*)=(i/2,i)$ indeed reproduce the free fermionic form of the partition function \eqref{GammaFF}.

Given the fermionic partition function \eqref{PF}, the one--loop potential is evaluated by summing over all inequivalent worldsheet tori via the modular invariant integral
\begin{equation}\label{PotentialTU}
    V_\text{one--loop}(T^{(i)},U^{(i)}) = -\frac{1}{2}\frac{M_s^4}{(2\pi)^4} \int_\mathcal{F}\frac{d^2\tau}{\tau_2^2}\, Z(\tau,\bar{\tau},T^{(i)},U^{(i)}),
\end{equation}
where in $ Z(\tau,\bar{\tau},T^{(i)},U^{(i)})$ we include the spacetime bosonic degrees of freedom arising from the worldsheet. In models with a $U(1)_A$, an additional contribution to the potential $V_D$ is generated as discussed in Section \ref{AnomU1}. Since this term is independent of the geometric moduli it provides a constant shift of the potential throughout moduli space. Hence we use this to write the vacuum energy as
\begin{equation}\label{VUplift}
    V_\text{total}=V_\text{one--loop}(T^{(i)},U^{(i)})+V_D,
\end{equation}
where $V_D$ is given in term of the trace of the anomalous $U(1)_A$ via \eqref{eq:VD}. 



In order to calculate the one--loop potential of our models we must be able to move away from this special point in the moduli space. The details of the translation of a free fermionic model into a $\Intr_2^N$ orbifold was developed in \cite{florakis} and used in \cite{fr1,fr2} to calculate one--loop potentials. Some details of this translation are given in Appendix \ref{appendix:translation}. 

The motivation for this procedure is that it enables us to move away from the free fermionic point in the moduli space. Although being fixed at this point allows for a generic analysis of many important features of a string model, for issues such as SUSY breaking and one--loop stability, analysis across the moduli space is required. 

In general, we can perturb away from the free fermionic point using marginal operators given by the Thirring Interactions \cite{ChangKumar} which take the form $J^i(z)\bJ^j(\bar{z})=:y^iw^i::\byy^j\byy^j:$. Writing these currents in bosonised form we identify the geometric moduli $J^i(z)\bJ^j(\bz)=\der X^i \overline{\der X}^j$. For symmetric $\ztwo$ orbifolds we parametrise the moduli space by a complex structure and K\"ahler modulus for each torus associated to the $\Gamma_{2,2}$ lattices: $(T^{(1)},U^{(1)})$, $(T^{(2)},U^{(2)})$ and $(T^{(3)},U^{(3)})$. These moduli span the familiar $SO(2,2)/SO(2)\times SO(2)$ moduli space of $\ztwo$ symmetric orbifolds. 

 Once we have installed this moduli dependence and followed the translation procedure we can calculate the one--loop potential numerically at specific moduli values using
\begin{equation}\label{PotentialTU}
    V_\text{one--loop}(T^{(i)},U^{(i)}) = -\frac{1}{2}\frac{M_s^4}{(2\pi)^4} \int_\mathcal{F}\frac{d^2\tau}{\tau_2^2}\, Z(\tau,\bar{\tau},T^{(i)},U^{(i)})
\end{equation}
where we integrate over the fundamental domain
\begin{equation}
    \mathcal{F} = \{\tau\in\mathbb{C}\;|\;|\tau|>1 \, , \, |\tau_1|<1/2\}.
\end{equation}
Calculating the one--loop potential is then an exercise in solving modular integrals. 

One important observation is that we have 6 complex moduli inside this integral rendering an analysis of the potential in all directions impractical. A logical approach used in \cite{fr1,fr2}, is to take the volume of the first torus associated to $\text{Im}(T^{(1)})=T_2$ and analyse the potential solely in this direction, with the other moduli all fixed at their values at the free fermionic point. This choice is somewhat arbitrary except in the case of an SSS breaking (discussed further in the section \ref{section:SUSYbreak}) where an internal shift in the first torus means that $T_2$ parameterises the SUSY--breaking and SUSY will be restored in the large volume limit $T_2\rightarrow \infty$. 

\subsection{Calculating $\Tr U(1)_A$ from the partition function}\label{section:TrU1PF}
In section \ref{AnomU1} we discussed how the FI $D$--term contributed at 2--loop to the potential of our model. Its magnitude depended on the trace of the fields charged under the $U(1)_A$ in our model, that propagate in the anomalous triangle diagram in four dimensions. There are two equivalent ways to calculate this trace. One way is to extract those states of the massless spectrum charged under $U(1)_A$ and add up their charges. This approach utilises free fermionic classification tools that are easily computerised. The details of this approach are given in Section \ref{section:chiralsecs}.
The second way to calculate $\Tr U(1)_A$ is directly from the partition function, which will be explained in this subsection. \\
\noindent
In order to perform this calculation it helps to rewrite the partition function (\ref{PF}) as follows
\begin{aline}\label{PF-ArgumentTheta}
    \Tilde{Z}=\,&\frac{1}{\eta^{11}\bar{\eta}^{22}} \, \frac{1}{2^{2}} \sum_{\substack{a,k\\b,l}} \;\frac{1}{2^{6}} \,\sum_{\substack{H_i\\G_i}} \;\frac{1}{2^{4}}\sum_{\substack{h_1,h_2,P_i\\g_1,g_2,Q_i}} (-1)^{a+b+P_1 Q_1 + P_2 Q_2+\Phi\smb{a&k&H_i&h_1&h_2&P_i}{b&l&G_i&g_1&g_2&Q_i}}\\[0.2cm]
    &\times \vth\smb{a}{b}(\nu)_{\psi^\mu} \vth\smb{a}{b}(\omega)_{\psi^\mu} \left( \frac{1}{\eta^2} \right) \vth\smb{a+h_1}{b+g_1}_{\chi^{12}} \vth\smb{a+h_2}{b+g_2}_{\chi^{34}} \vth\smb{a-h_1-h_2}{b-g_1-g_2}_{\chi^{56}}\\[0.3cm]
    &\times \Gamma^{(1)}_{2,2}\smbar{H_1&H_2}{G_1&G_2}{h_1}{g_1}(T^{(1)},U^{(1)}) \\[0.2cm]
    &\times \Gamma^{(2)}_{2,2}\smbar{H_3&H_4}{G_3&G_4}{h_2}{g_2}(T^{(2)},U^{(2)}) \\[0.2cm]
    &\times \Gamma^{(3)}_{2,2}\smbar{H_5&H_6}{G_5&G_6}{h_1+h_2}{g_1+g_2}(T^{(3)},U^{(3)}) \\[0.3cm]
     &\times \vthb\smb{k}{l}^5_{\bar{\psi}^{1-5}} \vthb\smb{k+h_1}{l+g_1}(u_1)_{\bar{\eta}^1} \vthb\smb{k+h_2}{l+g_2}(u_2)_{\bar{\eta}^2} \vthb\smb{k-h_1-h_2}{l-g_1-g_2}(u_3)_{\bar{\eta}^3} \vthb\smb{k+P_1}{l+Q_1}^4_{\bar{\phi}^{1-4}} \vthb\smb{k+P_2}{l+Q_2}^4_{\bar{\phi}^{5-8}}.
\end{aline}
where we have reinserted the orthogonal component of $\psi^\mu$ and its ghost contribution 
and the theta functions corresponding to the $\bar{\eta}^i$ fields acquire a non--zero argument $u_i$.
The total $U(1)_{i=1,2,3}$ traces will then be given as
\begin{equation}
    \Tr U(1)_i = \frac{\partial_\nu}{2 \pi} \frac{\partial_\omega}{2 \pi} \frac{\partial_{u_i}}{2\pi} \Tilde{Z}\Big|_{\nu,\omega,u_i = 0}
\end{equation}
where the derivatives $\partial_\nu \partial_\omega$ correspond to the correlator $\langle \psi \psi \psi \psi \rangle$, while $\partial_{u_i}$ corresponds to the $U(1)_i$ contribution of $\langle \bar{J}^i\rangle$ in (\ref{equation:VertexOperatorCalculation}) acting on the partition function.
The total anomalous $U(1)_A$ will be the combination of the three $U(1)_i$ according to (\ref{eqn:AnomalousU1}).

\section{Explicit vs spontaneous SUSY breaking}
\label{section:SUSYbreak}
The SUSY breaking $\mathcal{N}=1\rightarrow 0$ can happen in two ways: explicit or spontaneous (SSS) breaking. 
In the latter case, the GGSO phases can be set such that the gravitino acquires a mass and supersymmetry is broken spontaneously. The requirement is that the partition function, which will be non--zero in any point in the $T_2$ moduli space, vanishes when $T_2 \rightarrow \infty$ such that the potential in that limit vanishes and supersymmetry is restored. 
Spontaneous breaking certainly has attractive features compared with the explicit case. As discussed in various works \cite{SS2,SS1,SS3,CoCSuppression1,CoCSuppression2,CoCSuppression3,CoCSuppression4}, when accompanied by massless Bose--Fermi degeneracy at some point in the moduli space, $N_b^0=N_f^0$, we have the so--called `Super No--Scale' models in which the cosmological constant is exponentially suppressed according to \cite{Antoniadis90}
\beq 
\Lambda \propto (N^0_b - N^0_f) \frac{1}{T_2 ^2} + \mathcal{O}(e^{-c \sqrt{T_2}}),
\label{noscalesuppression}
\eeq
We note that without the $N_b^0=N_f^0$ condition, SSS models have polynomial, rather than exponential, suppression of their one--loop cosmological constant. 
However, as noted in \cite{fr1,fr2,fr3} such `super no--scale' conditions are merely necessary, not sufficient, conditions on the global structure of the
effective potential, which will be crucially dependent on the full mass tower of states, including the non level--matched ones around special self--dual points in moduli space.

Despite the attractive features of such `super no--scale' models, we note that the cosmological constant problem remains an issue, also in these models. 
Even in the case of spontaneous SUSY--breaking, there are sectors in the additive group $\Xi$ generated by the basis vectors that produce equal numbers of bosons and fermions \cite{aafs}. The states from these sectors do not reside in super--multiplets as supersymmetry is broken. As usual the respective bosonic and fermionic states arise from the generic sectors, {\it e.g.}  $\alpha\in\Xi$ and $S+\alpha\in \Xi$, and the bosonic 
and fermionic states from these sectors differ in some of their $U(1)$ charges, reflecting the fact that supersymmetry is broken. However, the phenomenological requirement still demands that the bosonic states from these sectors, that may, for example, correspond to the would be superpartners of the chiral generations, receive mass of the order of $1{\rm TeV}$. Generating this mass splitting between will produce a cosmological 
constant. Similarly, the other mass scales in the Standard Model, {\it e.g.} the QCD scale, will contribute to the cosmological constant. The cosmological constant problem is therefore much more profound, indicating a fundamental dichotomy between Quantum Field Theories expectations and gravitational observations, and it is naive to expect that the suppression observed in eq. (\ref{noscalesuppression}) can address the problem.  
%
We would ideally also consider higher loop contributions, {\it e.g.} the two--loop cosmological constant that should ideally be incorporated into this analysis. 
The fact is that the cosmological constant problem remains regardless of the SUSY--breaking mechanism. For this reason, we also explore some models with explicitly broken supersymmetry in our analysis. 
However, we also detail in the following how to identify the SSS models, which have their distinct phenomenological characteristics.

In the SSS SUSY--breaking the gravitino acquires a mass proportional to $\frac{1}{T_2}$, such that SUSY is restored at the border of the moduli space when  $T_2\rightarrow \infty$ and the partition function vanishes.
For this to happen, we require that in that limit the modular block in the partition function relating to the $\{\psi^\mu,\chi^{12},\chi^{34},\chi^{56}\}$ fermions gives rise to the Jacobi identity. To explore when this can happen, we set $\left( h_i,g_i \right)$ indices to zero since for all cases when SUSY is restored from $\mathcal{N}=0$ to $\mathcal{N}\rightarrow 1,2,4$, the partition function $Z$ will vanish.

The momenta in (\ref{PLPR}), through a redefinition of the summation variables $n_i \rightarrow n_i +m_i +H_i$ and fixing $T_1$, $U_1$ and  $U_2$ at the free fermionic point, can be written as follows
\begin{aline}
    \mathcal{P}_L =& \frac{1}{\sqrt{2 T_2 }} \left[ \left( m_1 + \frac{n_1}{2} + \frac{H_1}{2} - n_1 T_2 \right) + i \left( m_2 + \frac{n_2}{2} + \frac{H_2}{2} - n_2 T_2 \right) \right]  \\
    \mathcal{P}_R =& \frac{1}{\sqrt{2 T_2 }} \left[ \left( m_1 + \frac{n_1}{2} + \frac{H_1}{2} + n_1 T_2 \right) + i \left( m_2 + \frac{n_2}{2} + \frac{H_2}{2} + n_2 T_2 \right) \right].
\end{aline}
We can now observe that when $T_2\rightarrow \infty$ the only non--zero term in the moduli--dependent lattice sum (\ref{Z22TU}) is for $n_i =0$ and in this limit it contributes according to
\beq 
\Gamma^{(1)}_{2,2}\smbar{H_1&H_2}{G_1&G_2}{h_1}{g_1}(T_2^{(1)} \rightarrow \infty,T_{1*}^{(1)},U^{(1)}_*) \rightarrow \sum_{m_i\in\mathbb{Z}} 1,
\label{eq:LimitLattice}
\eeq 
where the sum, up to a normalization that will not affect our discussion, can be set to $1$. Then the partition function (\ref{PF}) will factorize as 
\begin{aline}\label{SS-PartitionFunction}
    Z& (T_2^{(1)}  \rightarrow \infty,T_{1*}^{(1)},U^{(1)}_*) =
    \,\frac{1}{\eta^{10}\bar{\eta}^{22}} \, \frac{1}{2^{12}} \sum_{\substack{k,H_i,P_i\\l,G_i,Q_i}} (-1)^{P_1 Q_1 + P_2 Q_2}\,
    \vthb\smb{k}{l}^8 \vthb\smb{k+P_1}{l+Q_1}^4 \vthb\smb{k+P_2}{l+Q_2}^4\\[0.2cm]
    &\times  \;\Gamma^{(2)}_{2,2}\smbar{H_3&H_4}{G_3&G_4}{0}{0}(T^{(2)} _*,U^{(2)} _*) \; \Gamma^{(3)}_{2,2}\smbar{H_5&H_6}{G_5&G_6}{0}{0}(T^{(3) }_*,U^{(3)} _*) \\[0.3cm]
    & \times \biggl( (-1)^{\Phi\smb{0&k&H_i&0&0&P_i}{0&l&G_i&0&0&Q_i}} \; \vth_3 ^4 +  (-1)^{1+\Phi\smb{1&k&H_i&0&0&P_i}{0&l&G_i&0&0&Q_i}} \;  \vth_2 ^4  +(-1)^{1+\Phi\smb{0&k&H_i&0&0&P_i}{1&l&G_i&0&0&Q_i}} \;  \vth_4 ^4 \biggl).
\end{aline}
The SSS condition requires 
\begin{equation}
    Z (T_2^{(1)}  \rightarrow \infty,T_{1*}^{(1)},U^{(1)}_*) = 0
    \label{eq:LimitPF}
\end{equation}
In order to vanish the phase has to satisfy
\begin{equation}    \sum_{\substack{H_1, H_2\\G_1, G_2}} (-1)^{\Phi\smb{0&k&H_i&0&0&P_i}{0&l&G_i&0&0&Q_i}} = \sum_{\substack{H_1, H_2\\G_1, G_2}} (-1)^{\Phi\smb{1&k&H_i&0&0&P_i}{0&l&G_i&0&0&Q_i}} = \sum_{\substack{H_1, H_2\\G_1, G_2}} (-1)^{\Phi\smb{0&k&H_i&0&0&P_i}{1&l&G_i&0&0&Q_i}} 
\label{eq:SS-Phase}
\end{equation}
which translates into a set of intricate conditions and relations between the GGSO phases. In Appendix \ref{appendix:SS-T-dualityConditions} we will show explicitly two examples, one satisfying the SSS conditions and the other with explicit SUSY breaking.

\subsection{T--duality}
\label{section:T-duality}
For our choice of models specified by the basis set (\ref{SO10basis}) T--duality is not always preserved. In particular, the symmetric shifts represented by the $\e{i}$ basis vectors may spoil the original $SL(2; \mathbb{Z})_T$ symmetry associated to the moduli $T$. We will now show how, for an SSS model, T--duality can be broken. As already specified in section \ref{PFSection}, we will vary $T_2$ only associated to the first 2--torus. 

The left and right momenta of the shifted lattice in (\ref{PLPR}) are left invariant under the following transformation
\begin{equation}
\label{eq:T-duality}
    T \rightarrow - \frac{1}{4 T} \;\;\; \leftrightarrow \;\;\; T_2 \rightarrow \frac{1}{4 T_2}.
\end{equation}
However, the phase in the lattice (\ref{Z22TU}) gets an additional term according to
\begin{equation}
\label{eq:T-dual-lattice}
e^{i\pi\sum_i((m_i+n_i+H_i)G_i} \rightarrow e^{i\pi\sum_i((m_i+n_i+H_i)G_i} \times e^{i\pi \left( H_1 G_1 + H_2 G_2 \right) },
\end{equation}
which will generally break T--duality. 
The same result can also be obtained following the discussion of section \ref{section:SUSYbreak}. In the $T_2 \rightarrow 0$ limit, with the other moduli fixed, the only non--zero contributions from the lattice (\ref{Z22TU}) are $1$, setting $m_i = n_i = H_i = 0$ in the lattice sum, $\sum_{m_i} 1$, setting $n_i = -2m_i, H_i =0$, and $e^{i\pi \left( H_1 G_1 + H_2 G_2 \right)}$ with $m_i = 0, n_i = -H_i$. The first two terms correspond, up to a normalization constant, to (\ref{eq:LimitLattice}) which for an SSS model give a vanishing contribution. Meanwhile, the third contribution generates an additional phase
\beq 
\Gamma^{(1)}_{2,2}\smbar{H_1&H_2}{G_1&G_2}{h_1}{g_1}(T_2^{(1)} \rightarrow 0,T_{1*}^{(1)},U^{(1)}_*) \rightarrow  e^{i\pi \left( H_1 G_1 + H_2 G_2 \right)},
\eeq 
In order to impose T--duality (\ref{eq:T-duality}), in addition to  (\ref{eq:LimitPF}) for the SSS condition, we must then require
\begin{equation}\label{T-duality-phases-1}
    Z (T_2^{(1)}  \rightarrow \infty,T_{1*}^{(1)},U^{(1)}_*) = Z (T_2^{(1)}  \rightarrow 0,T_{1*}^{(1)},U^{(1)}_*) = 0,
\end{equation}
with
\begin{aline}\label{T-duality-phases}
    Z& (T_2^{(1)}  \rightarrow 0,T_{1*}^{(1)},U^{(1)}_*) =
    \,\frac{1}{\eta^{10}\bar{\eta}^{22}} \, \frac{1}{2^{12}} \sum_{\substack{k,H_i,P_i\\l,G_i,Q_i}} (-1)^{P_1 Q_1 + P_2 Q_2 +H_1 G_1 + H_2 G_2 }\\[0.2cm] 
    &\times
    \vthb\smb{k}{l}^8 \vthb\smb{k+P_1}{l+Q_1}^4 \vthb\smb{k+P_2}{l+Q_2}^4
    \;\Gamma^{(2)}_{2,2}\smbar{H_3&H_4}{G_3&G_4}{0}{0}(T^{(2)} _*,U^{(2)} _*) \; \Gamma^{(3)}_{2,2}\smbar{H_5&H_6}{G_5&G_6}{0}{0}(T^{(3) }_*,U^{(3)} _*) \\[0.3cm]
    & \times \biggl( (-1)^{\Phi\smb{0&k&H_i&0&0&P_i}{0&l&G_i&0&0&Q_i}} \; \vth_3 ^4 +  (-1)^{1+\Phi\smb{1&k&H_i&0&0&P_i}{0&l&G_i&0&0&Q_i}} \;  \vth_2 ^4  +(-1)^{1+\Phi\smb{0&k&H_i&0&0&P_i}{1&l&G_i&0&0&Q_i}} \;  \vth_4 ^4 \biggl).
\end{aline}
As in (\ref{eq:SS-Phase}), T--duality now requires
\begin{aline}  
\label{eq:T-duality-condition}
\sum_{\substack{H_1, H_2\\G_1, G_2}} (-1)^{\Phi\smb{0&k&H_i&0&0&P_i}{0&l&G_i&0&0&Q_i}+H_1G_1 +H_2 G_2} & = \sum_{\substack{H_1, H_2\\G_1, G_2}} (-1)^{\Phi\smb{1&k&H_i&0&0&P_i}{0&l&G_i&0&0&Q_i}+H_1G_1 +H_2 G_2} \\
& = \sum_{\substack{H_1, H_2\\G_1, G_2}} (-1)^{\Phi\smb{0&k&H_i&0&0&P_i}{1&l&G_i&0&0&Q_i}+H_1G_1 +H_2 G_2},
\end{aline}
which again will correspond to specific constraints on the GGSO phases. 

\noindent
Models which satisfy (\ref{T-duality-phases-1}) will then exhibit a SSS SUSY breaking with unbroken T--duality and the one--loop potential will then have the following behavior
\begin{equation}
\label{eq:Potential-T-duality}
    V_\text{one--loop}(T_2) = V_\text{one--loop}\left(\frac{1}{4 T_2}\right)   
\end{equation}
In particular the extrema of the potential, either a maximum or a minimum, will lie 
at the self--dual point $T_2 = \frac{1}{2}$. 

We note that if instead of having $\e{i}$ in our basis, we had $\bm{T_j} = \bm{e_{2j -1}} +\bm{e_{2j}}$, $j=1,2,3$, as used in \cite{fr1,fr2}, the additional phase in (\ref{eq:T-dual-lattice}) is modified according to $e^{i\pi ( H_1 G_1 + H_2 G_2)} \rightarrow e^{i\pi ( H_1 G_1 + H_1 G_1)} = 1$. Thus for these models, with the indices $H_i, G_i \in \mathbb{Z}$, T--duality (\ref{eq:T-duality}) will always be satisfied. In Appendix \ref{appendix:SS-T-dualityConditions} we will show how for an SSS model the T--duality conditions can be implemented. 

\section{Tachyon projection} \label{section:tachs}
The tachyonic sectors in the models defined by the basis (\ref{SO10basis}) and their projection conditions are much the same as detailed in refs. \cite{so10tclass,PStclass}. The sectors and their mass levels are summarised in Table \ref{tachSectors}. 
\renewcommand{\arraystretch}{1.2}
\begin{table}[!ht]
\centering
\begin{tabular}{|c|c|c|}
\hline
Mass Level&	Vectorials & Spinorials \\
\hline
$(-1/2,-1/2)$&$\{\bar{\lambda}^m\}\ket{0}$& \ $\z{1},\ \z{2}$\\ 
\hline
$(-3/8,-3/8)$&$\{\bar{\lambda}^m\}\e{i}$&\ $\e{i}+\z{1}, \ \e{i}+\z{2}$\\
\hline
$(-1/4,-1/4)$& $\{\bar{\lambda}^m\}\e{i}+\bm{e_j}$& $\e{i}+\bm{e_j}+\z{1}, \ \e{i}+\bm{e_j}+\z{2}$\\
\hline
$(-1/8,-1/8)$&$\{\bar{\lambda}^m\}\e{i}+\bm{e_j}+\bm{e_k}$ & $\e{i}+\bm{e_j}+\bm{e_k}+\z{1}, \  \e{i}+\bm{e_j}+\bm{e_k}+\z{2}$\\
\hline
\end{tabular}
\caption{\label{tachSectors} \emph{Level-matched tachyonic sectors and their
    mass level, where $i\neq j \neq k=1,...,6$ and $\bar{\lambda}^m$ is any right-moving complex fermion with NS boundary condition for the relevant tachyonic sector.}}
\end{table}

In order to determine whether a sector survives the GGSO projections and remains in the spectrum we can construct a projector. For example, taking a sector with no oscillators, $\ket{\Bgb}$, the survival/projection condition is encapsulated in the generalised projector
\beq \label{GProj}
\mathbb{P}_{\Bgb}=\prod_{\bm{\xi}\in \Upsilon(\Bgb)}\frac{1}{2}\left( 1+\delta_{\Bgb} \CC{\Bgb}{\bm{\xi}}\right),
\eeq 
where 
\beq 
\delta_{\Bgb}=\begin{cases} +1 \ \ \text{if }  \beta(\psi^\mu)=0 \ \ \iff \ \ \text{sector is bosonic}\\
-1 \ \ \text{if }  \beta(\psi^\mu)=1 \ \ \iff \ \ \text{sector is fermionic}.
\end{cases}
\eeq 
The $\Upsilon(\Bgb)$ is defined as a minimal linearly independent set of vectors $\bm{\xi}$ such that $\bm{\xi} \cap \Bgb=\emptyset$. To check whether the sector $\Bgb$ is projected simply amounts to checking $\mathbb{P}_{\Bgb}=0$.

In the presence of a single right--moving oscillator $\bar{\lambda}$ with $\nu_f=\frac{1}{2}$, the generalised projector is modified to
\beq \label{GProjOScill}
\mathbb{P}_{\Bgb}=\prod_{\bm{\xi}\in \Upsilon(\Bgb)}\frac{1}{2}\left( 1+\delta_{\Bgb} \delta^{\bar{\lambda}}_{\bm{\xi}} \CC{\Bgb}{\bm{\xi}}\right),
\eeq 
such that 
\beq 
\delta^{\bar{\lambda}}_{\bm{\xi}}=\begin{cases}
+1 \ \ \text{if } \ \bar{\lambda} \in \bm{\xi} \\
-1 \ \ \text{if } \ \bar{\lambda}\notin \bm{\xi}
\end{cases}.
\eeq 

In order to build tachyon--free models we simply require that $\mathbb{P}_{\bm{t}}=0$ for all tachyonic sectors, $\bm{t}$. This requires defining the projection sets $\Upsilon(\bm{t})$ for each tachyonic sector of Table \ref{tachSectors}. For example, the tachyonic states from $\bm{z_1}$ have the projection set
\begin{align}
    \Upsilon(\z{1})&=\{\Sv,\e{1},\e{2},\e{3},\e{4},\e{5},\e{6},\x,\z{2}\}.
\end{align}
Checking for the absence of tachyonic sectors then amounts to checking the GGSO phases associated to such projection sets for all tachyonic sectors.
\section{Chiral sector analysis}
\label{section:chiralsecs}
Having explained how to get $\Tr U(1)_A$ directly from the partition function in subsection (\ref{section:TrU1PF}) we now explain how this can be calculated more efficiently by analysis of the sectors that produce massless states that can be charged under $U(1)_A$. 
This is equivalent to inspecting chiral sectors giving rise to states that are charged under the complex $\bget^{1,2,3}$ worldsheet fermion fields, where we recall that the charge of a free fermion is given by 
\beq 
Q(f)=\frac{1}{2}\alpha(f)+F(f),
\eeq 
and the action of the fermion number operator is
\beq 
F:\begin{cases}
f\ket{0}_{NS}=+1\\
f^*\ket{0}_{NS}=-1\\
\ket{+}=0\\
\ket{-}=-1
\end{cases},
\eeq 
where we write the two helicities of the degenerate Ramond vacuum as $\ket{\pm}$.

The first sectors we can identify with non--trivial chirality come from 
\begin{align}\label{spin16s}
    \begin{split}
        \bm{F}^1_{pqrs} =& \bm{S}+\bm{b_1}+p\bm{e_3}+q\bm{e_4}+r\bm{e_5}+s\bm{e_6}\\ 
        =& \{\psi^{\mu},\chi^{1,2},(1-p)y^3\bar{y}^3, pw^3\bar{w}^3,(1-q)y^4\bar{y}^4,qw^4\bar{w}^4 \\
        & (1-r)y^5\bar{y}^5,rw^5\bar{w}^5,(1-s)y^6\bar{y}^6,sw^6\bar{w}^6,\bar{\eta}^{1},\bar{\psi}^{1,\ldots,5}\}\\
        \bm{F}^2_{pqrs} =& \bm{S}+\bm{b_2}+p\bm{e_1}+q\bm{e_2}+r\bm{e_5}+s\bm{e_6}\\ 
        \bm{F}^3_{pqrs} =& \bm{S}+\bm{b_3}+p\bm{e_1}+q\bm{e_2}+r\bm{e_3}+s\bm{e_4},
    \end{split}
\end{align}
which we note are the sectors that generate the spinorial $\mathbf{16}/\overline{\mathbf{16}}$'s of our $SO(10)$ GUT, although we will not be interested in this aspect of our models in this work. 

Associated to these sectors are the projection sets
\begin{align}\label{FUpsilons}
    \begin{split}
        \Upsilon(\bm{F}^1_{pqrs})&=\{\bm{z_1},\bm{z_2},\bm{e_1},\bm{e_2}\}\\
        \Upsilon(\bm{F}^2_{pqrs})&=\{\bm{z_1},\bm{z_2},\bm{e_3},\bm{e_4}\}\\
        \Upsilon(\bm{F}^3_{pqrs})&=\{\bm{z_1},\bm{z_2},\bm{e_5},\bm{e_6}\},
    \end{split}
\end{align}
which are used to determine whether a sector remains in the Hilbert space of the model, just as explained for the tachyonic sectors in the previous section. 

Once we have checked the survival of a particular sector, we can then determine the chirality of the $\bget^{1,2,3}$ for the resultant states through the chirality projections defined for the three orbifold planes as follows
\begin{align}\label{ChProjs}
\begin{split}
&\chi(\bm{F}^{1}_{pqrs})=\text{ch}(\bar{\eta}^{1})=-\text{ch}(\psi^\mu) \ \CC{\bm{F}^{1}_{pqrs}}{\Sv+\bv{2}+\x+(1-r)\e{5}+(1-s)\e{6}}^*\\
&\chi(\bm{F}^{2}_{pqrs})=\text{ch}(\bar{\eta}^{2})=-\text{ch}(\psi^\mu) \ \CC{\bm{F}^{2}_{pqrs}}{\Sv+\bv{1}+\x+(1-r)\e{5}+(1-s)\e{6}}^*\\
&\chi(\bm{F}^{3}_{pqrs})=\text{ch}(\bar{\eta}^{3})=-\text{ch}(\psi^\mu) \  \CC{\bm{F}^{3}_{pqrs}}{\Sv+\bv{1}+\x+(1-r)\e{3}+(1-s)\e{4}}^*.
\end{split}
\end{align}
Without loss of generality we can choose ch$(\psi^\mu)=\ket{+}$ since the CPT--conjugates are necessarily present with the opposite chirality choice. This then fully determines the charges under the $U(1)_{1,2,3}$ for these sectors. 

Along with these 3 groups of 16 sectors we have a further 12 such groups $\bm{F}^{4,5,6}_{pqrs},\bm{F}^{7,8,9}_{pqrs}$ and $\bm{V}^{1,2,3}_{pqrs}$ but we relegate the details of how to extract their charge contributions to Appendix \ref{appendix:ChiralSec}. As with the analysis of the tachyonic sectors, the projection conditions that determine the overall $\Tr U(1)_A$ can then be computerised to allow for efficient scans of large spaces of different GGSO phase configurations. The results from such a scan is presented in the next section along with the results of our analysis of the $D$--term uplifted potentials.

As mentioned above, in this work we will not consider extra phenomenological issues in our models such as the number of spinorial $\mathbf{16}$'s or vectorial $\mathbf{10}$'s classified in previous works for these symmetric $\ztwo$ models \cite{slmclass,class1,fknr,fkr1,acfkr2,su4,su62,frs,lrsclass}. One may wonder about the relationship between such characteristics and the value of $\Tr U(1)_A$. Since viable $SO(10)$ phenomenology requires a condition such as $N_{\mathbf{16}}-N_{\mathbf{\overline{16}}}\geq 6$ we could expect some relationship between the trace values and the presence of an appropriate number of these representations. However, since  $\bm{F}^{4,5,6,7,8,9}_{pqrs}$ are hidden sectors and the vectorials $\bm{V}^{1,2,3}_{pqrs}$ would only be constrained by the presence of at least vectorial $\mathbf{10}$ we don't expect there to be any significant change to whether we can find $D$--term uplifted models once we incorporate such phenomenological considerations into the analysis.

\section{Results}\label{results}
The methodology we use for the extraction of $D$--term uplifted models defined through the basis (\ref{SO10basis}) follows the 5 step procedure:
\begin{enumerate}
    \item Extract $\mathcal{N}=0$ models, by checking whether eq (\ref{eqn:ProjectorGravitino}) and /or eq. (\ref{eqn:SUSYChiral}) are violated. Subsequently, we check that the models are free from physical tachyons by checking the projection conditions outlined in section (\ref{section:tachs}).
    \item Compute the values of $\Tr U(1)_A$ using the efficient analysis of chiral sectors explained in section \ref{section:chiralsecs} and appendix \ref{appendix:ChiralSec} for these $\mathcal{N}=0$ tachyon--free models.
    \item Extract out those models with larger $\Tr U(1)_A$ values and satisfying the SSS SUSY breaking conditions discussed in section \ref{section:SUSYbreak}.
    \item Perform the numerical analysis of the one--loop potentials and check for an uplift from AdS to dS for these SSS models with large $\Tr U(1)_A$.
    \item The models with explicit SUSY breaking but large $\Tr U(1)_A$ can then be analysed and checked for an uplift. 
\end{enumerate}
We note that the key bottleneck in this methodology is performing the numerical analysis of the one--loop potential integral(s). Depending on how many points with respect to $T_2$ are evaluated, the number of models we can analyse the potentials for in reasonable computing time is approximately only $\mathcal{O}(10^3)$. This helps to motivate the 5 step procedure above that seeks to maximise the probability we find an uplifted model, with or without SSS breaking.  

\subsection{Distribution of $\Tr U(1)_A$ for $\mathcal{N}=0$ models}
\begin{figure}[!ht]
\centering
\includegraphics[width=0.98\linewidth]{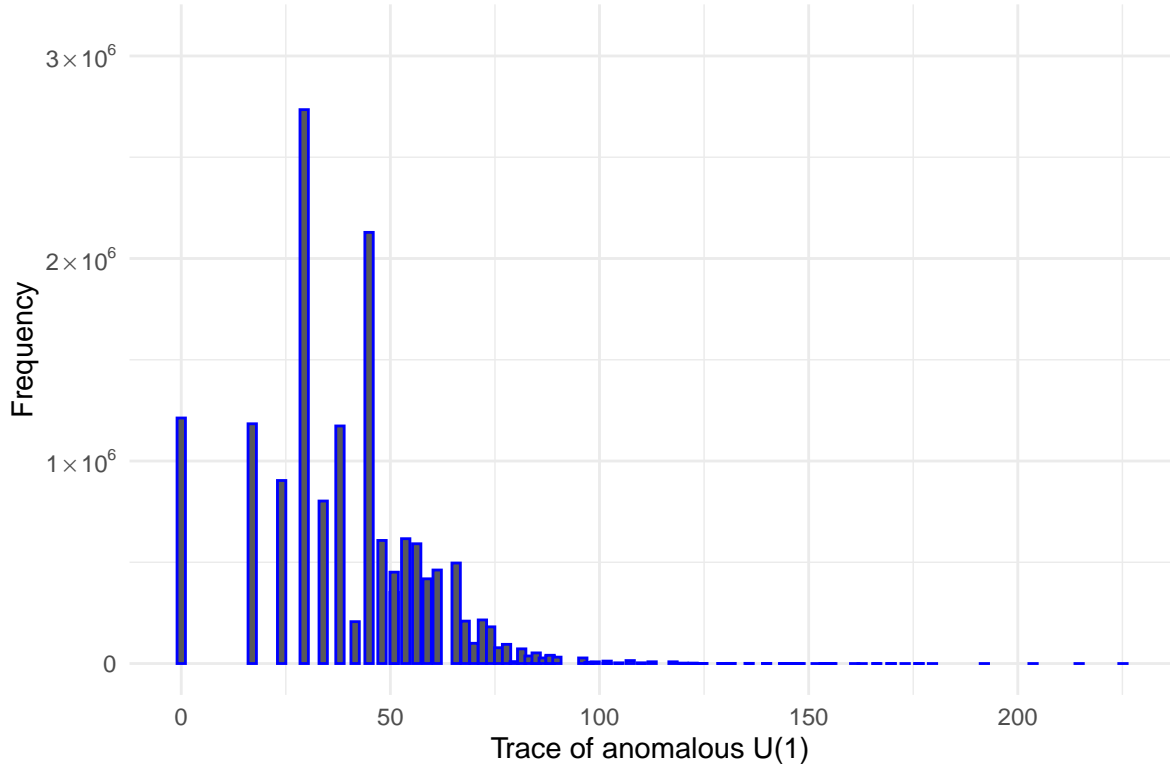}
\caption{\emph{The distribution of Tr $U(1)_A$}}
\label{U1ADist}
\end{figure}

For our purpose of finding an uplifted model it was sufficient in step 1. to take a random scan of $10^9$ $\mathcal{N}=0$ GGSO configurations and checking them for the absence of physical tachyons using the conditions explained in section \ref{section:tachs}. This scan resulted in $\sim 1.6 \times 10^7$ tachyon--free models. Following step 2. of the methodology, we then calculated the values of $\Tr U(1)_A$ for these models. The results for the distribution of these values are shown in Figure \ref{U1ADist}. 

\subsection{Example Scherk--Schwarz models}

Once the $\Tr U(1)_A$'s for tachyon--free models were founds, we moved to step 3. to begin the analysis of the one--loop potential starting with those models that had a larger value for $\Tr U(1)_A$ and also satisfied the SSS condition derived in section \ref{section:SUSYbreak}. Then we perform step 4. to search for the $D$--term uplift for these larger $\Tr U(1)_A$ SSS models. 

\begin{figure}[!ht]
\centering
\includegraphics[width=0.85\linewidth]{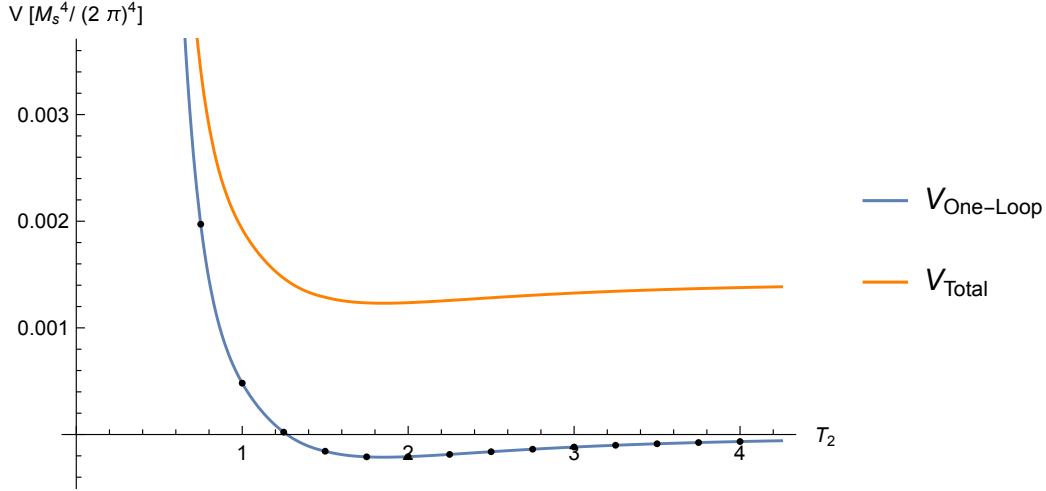}
\caption{\emph{One-loop Potential of SSS Example Model with Uplift}}
\label{figure:SSUpliftedPot}
\end{figure}

In a scan of approximately $10^3$ such models we did indeed find a single model that exhibited the desired uplift. This model is defined through the following set of GGSO phases 
{\begin{equation}
\small
\CC{\bm{v_i}}{\bm{v_j}}= 
\begin{blockarray}{cccccccccccccc}
&\mathbf{1}& \Sv & \e{1}&\e{2}&\e{3}&\e{4}&\e{5}&\e{6}&\bv{1} & \bv{2} &\z{1} & \z{2} \\
\begin{block}{c(rrrrrrrrrrrrr)}
\mathbf{1}&-1&-1& -1& 1&-1& 1&-1& 1& -1& -1&-1& 1&\ \\ 
\Sv       &-1&-1&1&1&-1&-1&-1&-1& 1&-1& 1& -1&\ \\
\e{1}     & -1&1&1&-1&-1& -1& 1& 1& 1& -1&-1&-1&\ \\ 
\e{2}     & 1&1&-1&-1&1&1&1&1& 1&-1& -1& 1&\ \\ 
\e{3}     &-1&-1&-1&1& 1&1&1&-1&1&1&1&1&\ \\
\e{4}     & 1&-1& -1&1&1&-1& 1& -1&1&1&1&1&\ \\
\e{5}     &-1&-1& 1&1&1& 1& 1&1& 1&1&1& 1&\ \\ 
\e{6}     & 1&-1& 1&1&-1& -1&1&-1& -1&-1&1&1&\ \\ 
\bv{1}    & -1&-1& 1& 1&1&1& 1& -1& -1& -1& -1&1&\ \\ 
\bv{2}    & -1& 1& -1&-1&1&1&1&-1& -1& -1& -1&1&\ \\ 
\z{1}     &-1&1&-1& -1&1&1&1&1& -1& -1&-1&1&\ \\
\z{2}     & 1& -1&-1& 1&1&1& 1&1&1&1&1& 1&\ \\
\end{block}
\end{blockarray}
\label{SSmodelGGSO}
\end{equation}}
and has a one--loop cosmological constant value of $\Lambda=-0.000212496$ at the free fermionic point and 
$\Tr U(1)_A =72 \sqrt{2}$,
with a FI contribution of $0.00144365$. Performing the numerical analysis allowed us to graph the potential and demonstrate its uplift, as shown in Figure \ref{figure:SSUpliftedPot}.

\begin{figure}[H]
\centering
\begin{minipage}{.45\textwidth}
  \centering
\includegraphics[width=1.\linewidth]{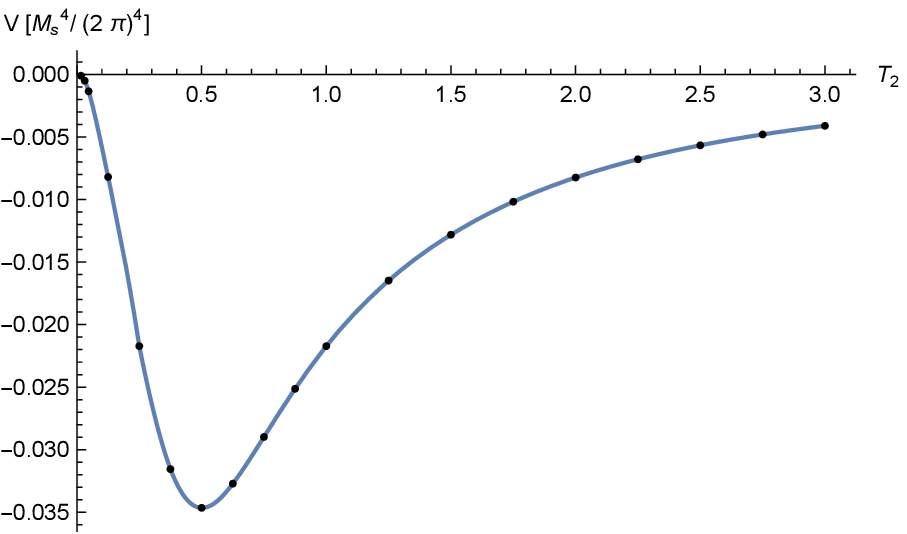}
  \captionof{figure}{\centering \emph{One-loop Scherk-Schwarz Potential with local minimum and unbroken T-duality}}
  \label{fig:30}
\end{minipage}%
\hspace{0.8cm}
\begin{minipage}{.45\textwidth}
  \centering
 \includegraphics[width=1.\linewidth]{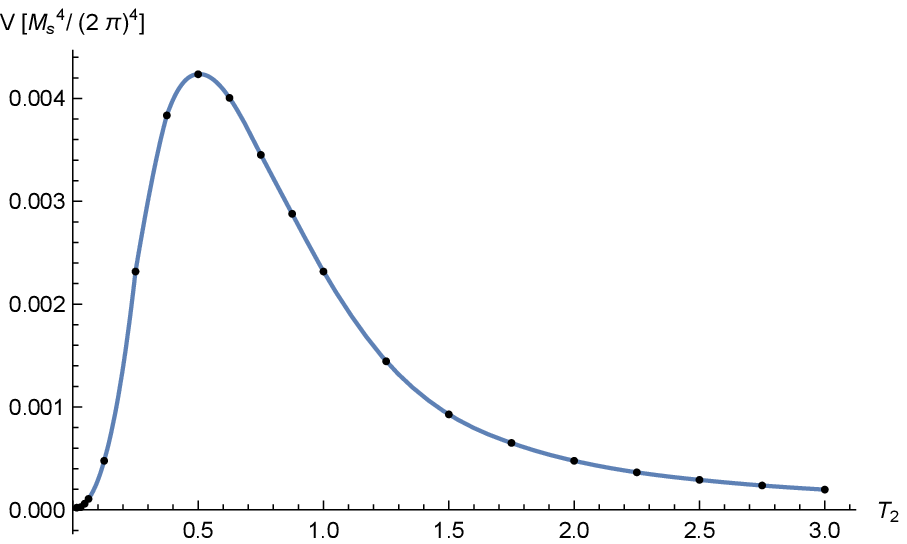}
  \captionof{figure}{\centering \emph{One-loop Scherk-Schwarz Potential with local maximum and unbroken T-duality}}
  \label{fig:31}
\end{minipage}
\end{figure}

\begin{figure}[H]
\centering
\begin{minipage}{.45\textwidth}
  \centering
 \includegraphics[width=1.\linewidth]{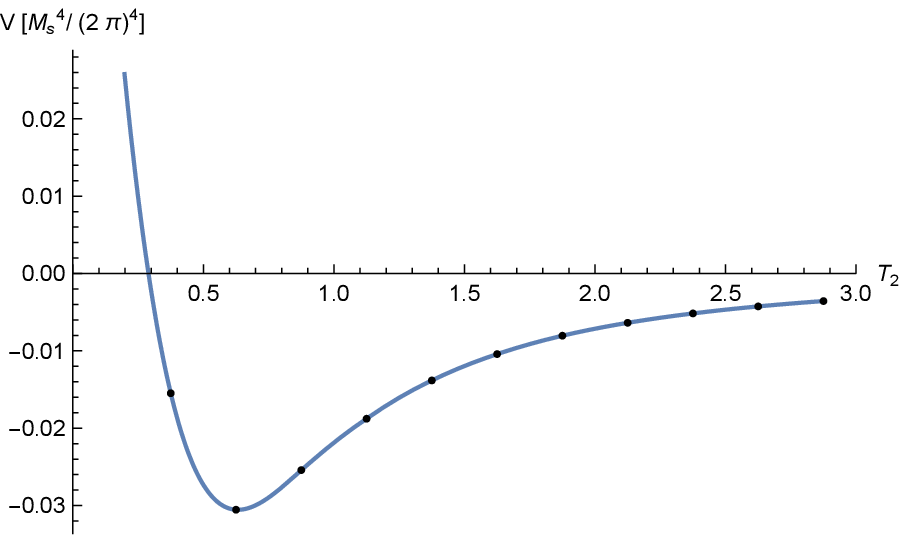}
  \captionof{figure}{\centering \emph{One-loop Scherk-Schwarz Potential with local minimum and broken T-duality}}
  \label{fig:7}
\end{minipage}
\hspace{0.8cm}
\begin{minipage}{.45\textwidth}
  \centering
\includegraphics[width=1.\linewidth]{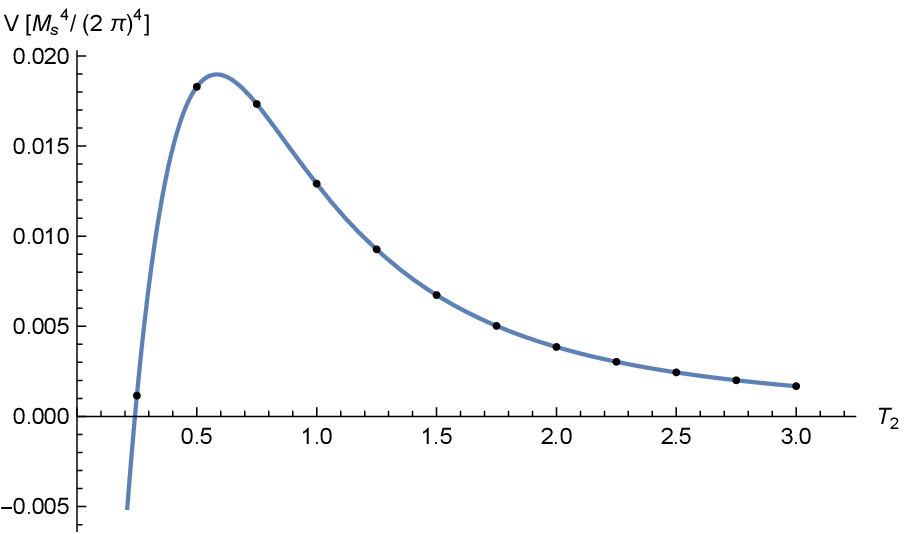}
  \captionof{figure}{\centering \emph{One-loop Scherk-Schwarz Potential with local maximum and broken T-duality}}
  \label{fig:29}
\end{minipage}%
\end{figure}

\begin{figure}[H]
\centering
\begin{minipage}{.45\textwidth}
  \centering
\includegraphics[width=1.\linewidth]{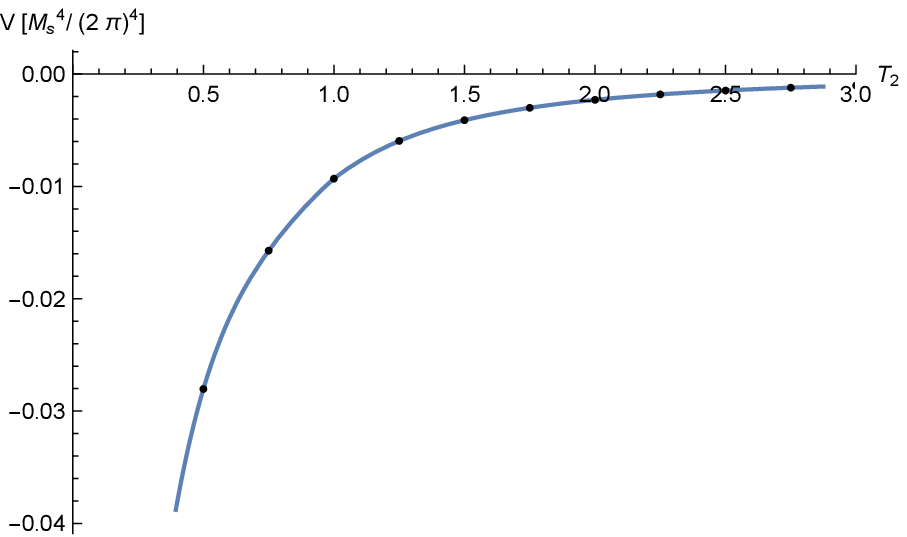}
  \captionof{figure}{\centering \emph{One-loop Scherk-Schwarz Potential without any extreme point and broken T-duality}}
  \label{fig:2_2}
\end{minipage}%
\hspace{0.8cm}
\begin{minipage}{.45\textwidth}
  \centering
 \includegraphics[width=1.\linewidth]{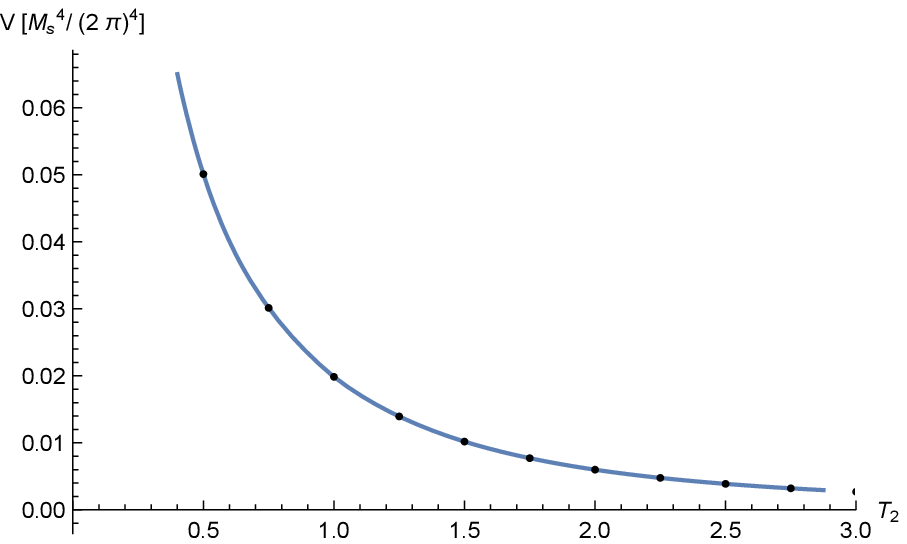}
  \captionof{figure}{\centering \emph{One-loop Scherk-Schwarz Potential without any extreme point and broken T-duality}}
  \label{fig:2_3}
\end{minipage}
\end{figure}

Through an analysis of $\mathcal{O}(10^3)$ GGSO configurations, we also evaluated the one--loop potential shapes for SSS broken models. We summarise these possibilities in Figures \ref{fig:30}-
\ref{fig:2_3}. We find two examples of SSS models with unbroken T--duality, see Figures \ref{fig:30} and \ref{fig:31}, and two with broken T--duality, see Figures \ref{fig:7} and \ref{fig:29}, where the difference can be clearly seen from the behavior of the potential when the modulus approaches zero. Figures \ref{fig:2_2} and \ref{fig:2_3} instead show two SSS models with broken T--duality and with no minima or maxima.

The models we are especially interested in for our purpose are the ones with a negative minima. In these cases the additional $D$--term contribution (\ref{eq:VD}) typically is not sufficient to uplift the minima. It is only in very rare, finely tuned, examples that this uplift is observed. Finally, we note that although we projected out physical
tachyons at the free fermionic point, in general there may be tachyons at other points in
the moduli space. However, through analysis across the range of the modulus $T_2$ we find
that none of our potentials suffer tachyonic instabilities. 



\subsection{Explicitly broken model examples}

\begin{figure}[!ht]
\centering
\includegraphics[width=0.85\linewidth]{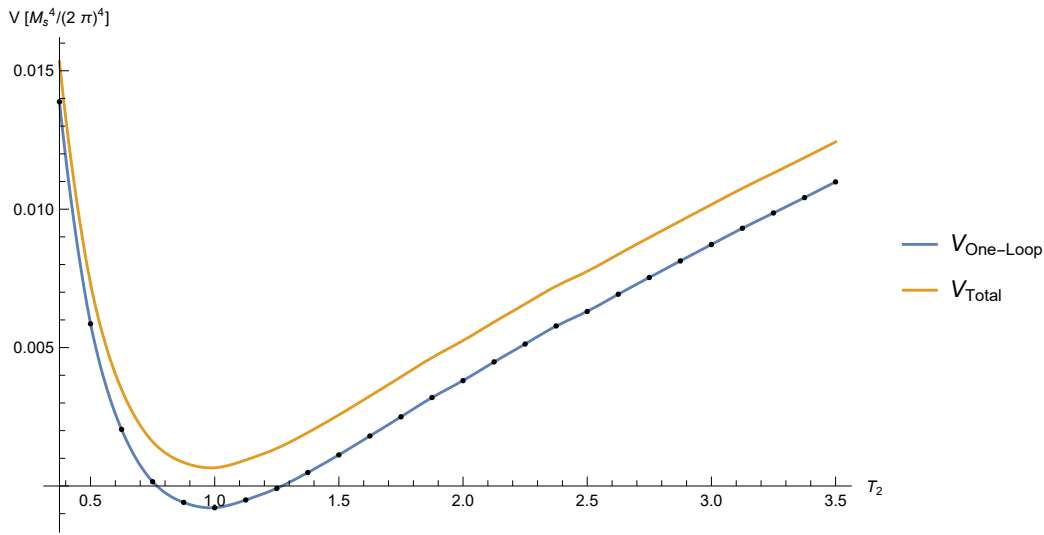}
\caption{\emph{One-loop Potential of Example Model with explicitly broken SUSY with Uplift}}
\label{figure:UpliftedPot}
\end{figure}

Performing a similar scan of large $\Tr U(1)_A$ models but with explicit breaking in step 5., we also find models in which the FI $D$--term uplifts the one--loop potential. An example of such a model with explicit SUSY--breaking is given by the GGSO phases
{\begin{equation}
\small
\CC{\bm{v_i}}{\bm{v_j}}= 
\begin{blockarray}{cccccccccccccc}
&\mathbf{1}& \Sv & \e{1}&\e{2}&\e{3}&\e{4}&\e{5}&\e{6}&\bv{1} & \bv{2} &\z{1} & \z{2} \\
\begin{block}{c(rrrrrrrrrrrrr)}
\mathbf{1}&-1&-1& 1& 1&-1& 1&-1& 1& 1& 1&-1& 1&\ \\ 
\Sv       &-1&-1&-1&-1&-1&-1&-1&-1& 1&-1&-1& 1&\ \\
\e{1}     & 1&-1&-1&-1&-1& 1& 1& 1& 1& 1&-1&-1&\ \\ 
\e{2}     & 1&-1&-1&-1&-1&-1&-1&-1& 1&-1& 1& 1&\ \\ 
\e{3}     &-1&-1&-1&-1& 1&-1&-1&-1&-1&-1&-1&-1&\ \\
\e{4}     & 1&-1& 1&-1&-1&-1& 1& 1&-1&-1&-1&-1&\ \\
\e{5}     &-1&-1& 1&-1&-1& 1& 1&-1& 1&-1&-1& 1&\ \\ 
\e{6}     & 1&-1& 1&-1&-1& 1&-1&-1& 1&-1&-1&-1&\ \\ 
\bv{1}    & 1&-1& 1& 1&-1&-1& 1& 1& 1& 1& 1&-1&\ \\ 
\bv{2}    & 1& 1& 1&-1&-1&-1&-1&-1& 1& 1& 1&-1&\ \\ 
\z{1}     &-1&-1&-1& 1&-1&-1&-1&-1& 1& 1&-1&-1&\ \\
\z{2}     & 1& 1&-1& 1&-1&-1& 1&-1&-1&-1&-1& 1&\ \\
\end{block}
\end{blockarray}
\label{GGSOModel2}
\end{equation}}
\noindent
corresponding to a model with one--loop cosmological constant $\Lambda=-0.000785598$ and 
$\Tr U(1)_A =72 \sqrt{2}$
that generates a FI contribution of $0.00144365$ to the one--loop potential, ensuring a positive minimum as depicted in Figure \ref{figure:UpliftedPot}.
\\

\begin{figure}[H]
\centering
\begin{minipage}{.45\textwidth}
  \centering
\includegraphics[width=1.\linewidth]{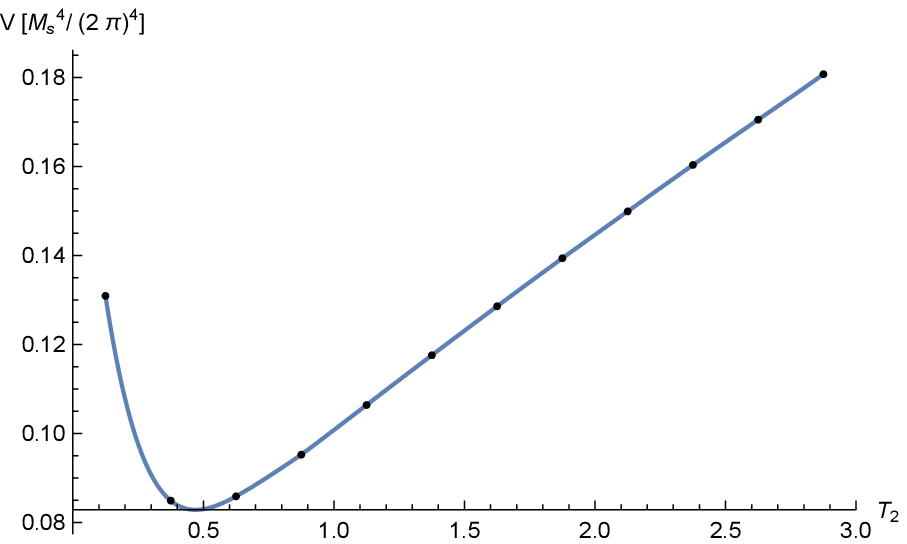}
  \captionof{figure}{\centering \emph{One-loop Potential with explicitly broken SUSY and local minimum}}
  \label{fig:5}
\end{minipage}%
\hspace{0.8cm}
\begin{minipage}{.45\textwidth}
  \centering
 \includegraphics[width=1.\linewidth]{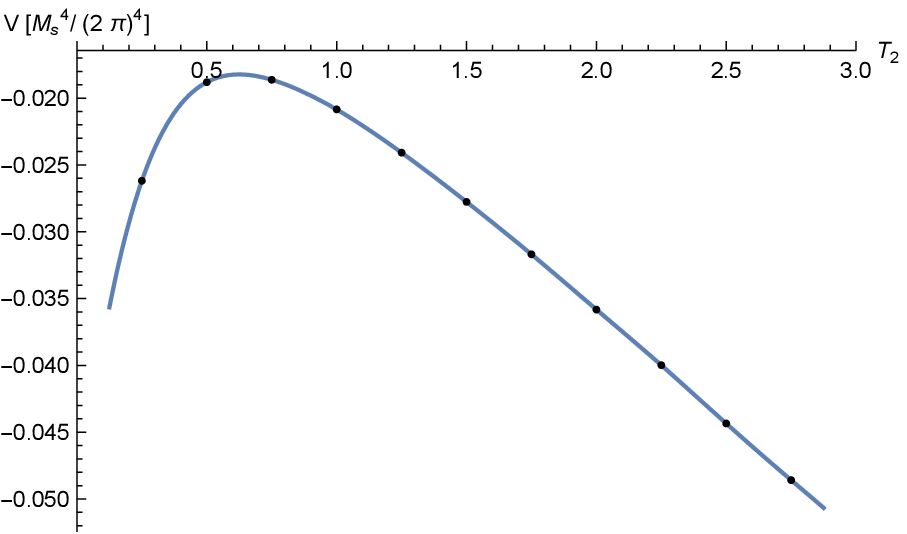}
  \captionof{figure}{\centering \emph{One-loop Potential with explicitly broken SUSY and local maximum}}
  \label{fig:27}
\end{minipage}
\end{figure}

\begin{figure}[H]
\centering
\begin{minipage}{.45\textwidth}
  \centering
\includegraphics[width=1.\linewidth]{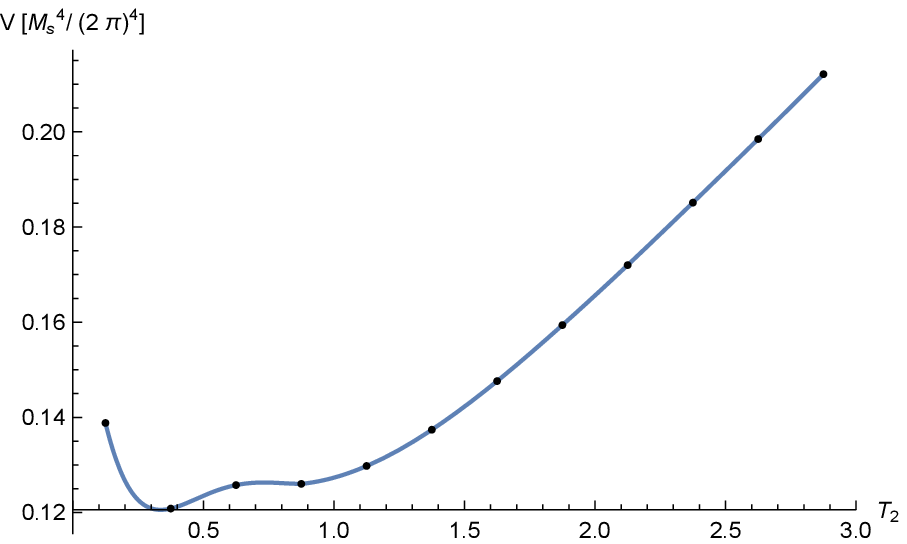}
  \captionof{figure}{\centering \emph{One-loop Potential with explicitly broken SUSY and local minima and maximum}}
  \label{fig:26}
\end{minipage}%
\hspace{0.8cm}
\begin{minipage}{.45\textwidth}
  \centering
 \includegraphics[width=1.\linewidth]{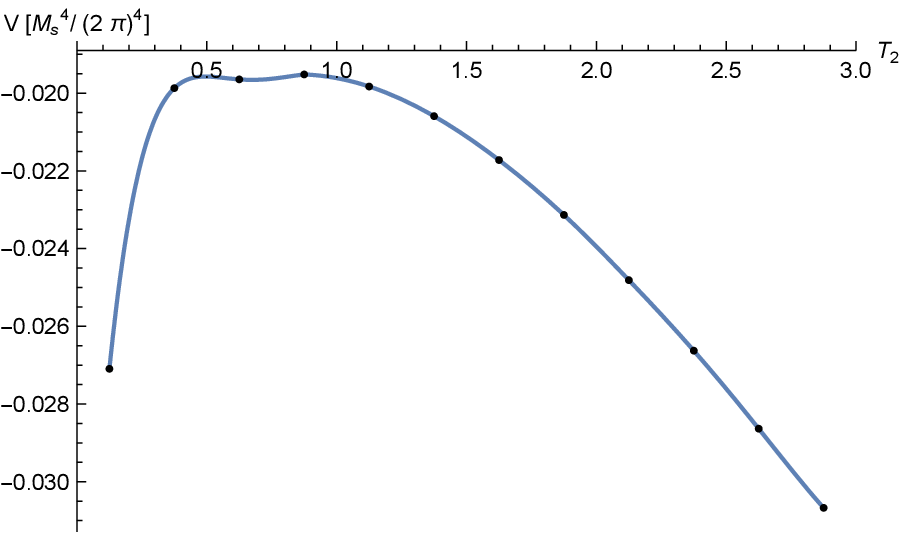}
  \captionof{figure}{\centering \emph{One-loop Potential with explicitly broken SUSY and local minimum and maxima}}
  \label{fig:22}
\end{minipage}
\end{figure}

\begin{figure}[H]
\centering
\begin{minipage}{.45\textwidth}
  \centering
\includegraphics[width=1.\linewidth]{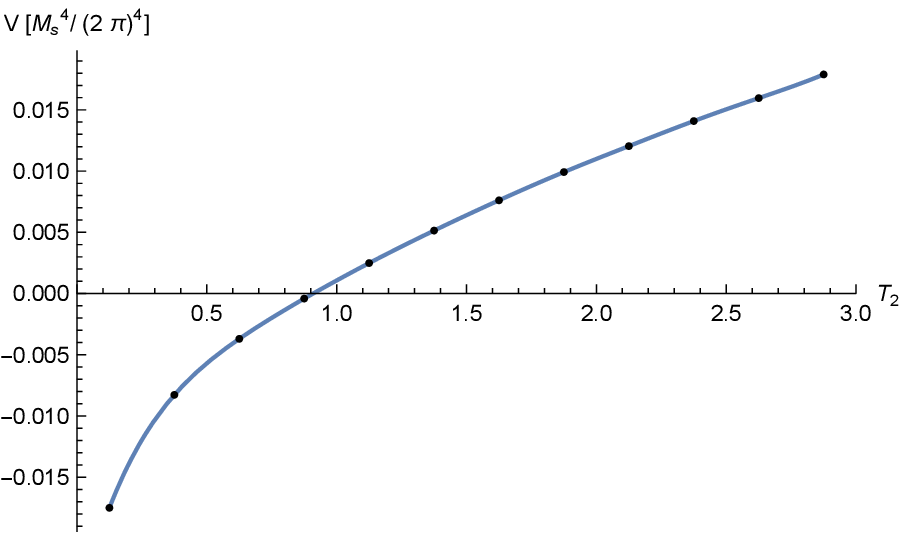}
  \captionof{figure}{\centering \emph{One-loop Potential with explicitly broken SUSY without any extreme point}}
  \label{fig:2}
\end{minipage}%
\hspace{0.8cm}
\begin{minipage}{.45\textwidth}
  \centering
 \includegraphics[width=1.\linewidth]{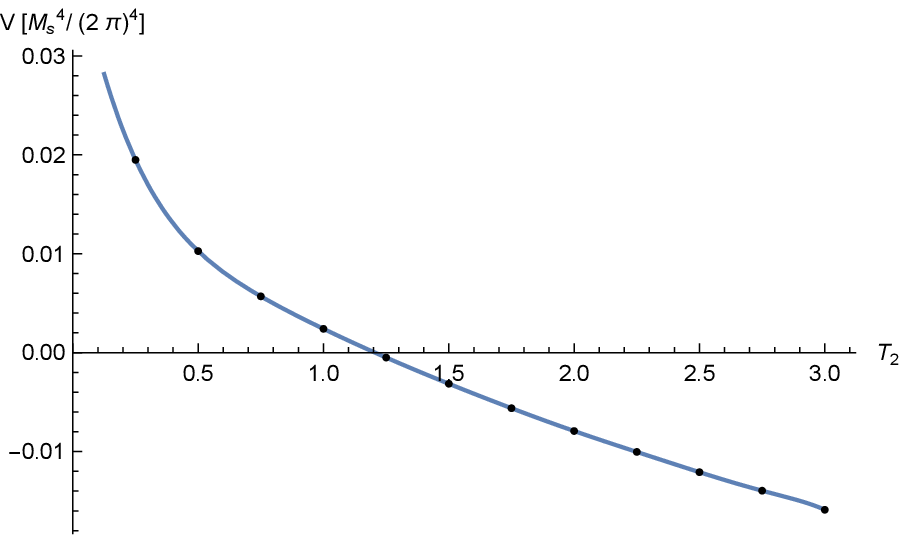}
  \captionof{figure}{\centering \emph{One-loop Potential with explicitly broken SUSY without any extreme point}}
  \label{fig:28}
\end{minipage}
\end{figure}

Also in this case, through analysis of $\mathcal{O}(10^3)$ models with explicitly broken SUSY, we find only certain possibilities for the shapes of the potential. We summarise these possibilities in Figures \ref{fig:5}-
\ref{fig:28}. In all these graphs, when the moduli $T_2 \rightarrow \infty$, the potential diverges so SUSY is broken explicitly. \\
In particular in Figures \ref{fig:5} and \ref{fig:26} the minima is positive while in Figure \ref{fig:27} only a negative maxima is present. In Figures \ref{fig:22} and \ref{fig:28} instead there are no extremal points at all. These are not the cases we are interested in. Only the shapes of the potential as in Figure \ref{fig:22} present a negative minima required for the uplift. 

\section{Discussion \& Conclusions}\label{conclusions} 

One important, and generic, issue with non--SUSY theories is the issue of a non--vanishing dilaton tadpole within such models with a non--zero cosmological constant. This means that the string equations of motion are not satisfied for our theory on a Minkowskian 4D spacetime with a constant dilaton. If we want to find the true perturbative quantum vacuum, we would need to solve
the string equations of motion to all loop levels. However, for non--SUSY 
strings we generically lose computational control at higher orders in the string loop expansion.

We note that in that respect the analysis performed in this paper is rather heuristic. The dilaton VEV was fixed by hand and most of the moduli are set at the free fermionic point but are not fixed, {\it i.e.} they can be varied away from the free fermionic point. Similarly, the two--loop contribution to the vacuum energy was adopted from the SUSY cases and not carried our explicitly. We remark that while our arguments here are, in these respects, not conclusive, the analysis can be made more rigorous by adopting the Kiritsis--Kounnas modular invariant regularisation method of one--loop string amplitudes \cite{KKregular}. In that scheme, the four dimensional spacetime curvature is used as a infrared regulator, and is achieved by compactifying the four 
dimensional spacetime on a Wess--Zumino--Witten current algebra sigma model. The simplest solution 
is obtained with the conformal theory $W_k^{(4)}= SU(2)_k\times U(1)$, which has the asymptotic (large $k$)
geometry $S^3\times R^+$. The heterotic string constructions with a mass gap for the massless spectrum
is then constructed by substituting the worldsheet coordinate and the spin--fields of the 
uncompactified Minkowski spacetime conformal and superconformal blocks of appropriate worldsheet 
supersymmetry and central charge. It should be noted that while this regularisation methodology was developed for 
vacua with $\mathcal{N}=1$ spacetime supersymmetry, its adaptation to non--SUSY string 
vacua should be possible. This scheme therefore provide the tools to construct non--SUSY
string solutions with regulated infrared divergences. Similarly, while perhaps presenting a substantial 
technical challenge, the two--loop costribution to the vacuum energy can in principle be calculated 
directly in string theory \cite{AS, CoCSuppression4}. Thus, the somewhat heuristic arguments made here
can, in principle, be put on firmer grounds. 

We also remark that the question of moduli stabilisation can also be partially addressed directly in the 
heterotic string worldsheet constructions. String theory allows for asymmetric boundary conditions 
of the left-- and right--moving worldsheet internal fermions, which the corresponding bosonic
representation entail an asymmetric action on the internal coordinates of the compactified
six dimensional space. As a consequence, some or even all of the moduli fields that parameterise 
the properties of the internal manifold can be projected out \cite{moduli} and be frozen at a specific value, 
typically at the self--dual point in the moduli space. In that case, the space of unfixed 
moduli is substantially reduced. One can envision that stabilisation of the remaining unfixed moduli 
and extraction of the global minimum is possible. The question whether it is possible to obtain a model
with positive vacuum energy and fixed moduli with or without a $D$--term uplift may be further 
investigated. 

In this paper we explored the contribution of the would--be FI $D$--term to 
the vacuum energy in non--SUSY heterotic string vacua. This contribution can 
uplift the vacuum energy from a negative to positive value and give rise to a positive constant 
cosmological background. We distinguished in our analysis between string vacua with explicit
SUSY breaking versus vacua in which SUSY is broken spontaneously by the 
Scherk--Schwarz mechanism. We found that while rare, a $D$--term uplift to a positive cosmological
constant might indeed be possible.



\section*{Acknowledgments}

AEF would like to that the CERN theory division for hospitality and support. 
The work of ARDA is supported in part by EPSRC grant EP/T517975/1.
The work of BP is supported by EPSRC grant EP/W522399/1.

\appendix
\section{Translation of Fermionic Partition Function}\label{appendix:translation}


The goal of this appendix is to find a one--to--one correspondence between the partition functions written in the free fermionic construction, (\ref{FFPF}), and in the orbifold language. An outline of this procedure can be found in \cite{florakis} and is also presented in \cite{Matyas:2023vgm}. 

The aim of this procedure is to find an equality of the form
\begin{equation}\label{PFTransEq}
   Z_F= \frac{1}{2^N}\sum_{\bm{\alpha},\bm{\beta}}\CC{\bm{\alpha}}{\bm{\beta}} Z\sqb{\bm{\alpha}}{\bm{\beta}} = \sum_{\substack{a,k,\cdots\\b,l,\cdots}} (-1)^{\Psi\smb{a&k&\cdots}{b&l&\cdots}} \;\,Z\sqb{a&k&\cdots}{b&l&\cdots},
\end{equation}
where the product over the fermions is now implicit and contained within $Z\smb{\bm{\alpha}}{\bm{\beta}}$. The right--hand side requires further comments. The term $Z\smb{a&k&\cdots}{b&l&\cdots}$ represents the theta functions in terms of the summation indices $a,b,k,l,\dots$. The phase $(-1)^{\Psi\smb{a&k&\cdots}{b&l&\cdots}}$ is the analogue of the GGSO phase in this new formulation.

\subsection*{The Modular Invariant Phase}

To see all possible choices of indices, which in turn fix the form of $Z\smb{a&k&\cdots}{b&l&\cdots}$, we note that to represent a partition function of a model with $N$ basis vectors requires the use of $N$ summation indices. This can be seen by matching the number of terms on each side of (\ref{PFTransEq}). Thus, the translation of the form of the partition function 
is uniquely determined by the choice of a change of basis matrix, $S$, which encodes the correspondence between the basis vectors and the summation indices. For our choice of models specified by the basis vectors ($\ref{SO10basis}$) and the summation indices in the partition function as in (\ref{PF}), the $S$ matrix is given by
{\begin{equation}
\small
S = 
\begin{blockarray}{cccccccccccccc}
\begin{block}{c(rrrrrrrrrrrrr)}
&\ 1&1& 1& 1&1& 1&1& 1& 0& 0&0& 0 \ \\ 
&\ 1&0&0&0&0&0&0&0& 0&0& 0& 0 \ \\
&\ 0&0&1&0&0& 0& 0& 0& 0& 0&0&0 \ \\ 
&\ 0&0&0&1&0&0&0&0& 0&0& 0& 0 \ \\ 
&\ 0&0&0&0& 1&0&0&0&0&0&0&0 \ \\
&\ 0&0& 0&0&0&1& 0& 0&0&0&0&0 \ \\
&\ 0&0& 0&0&0& 0& 1&0& 0&0&0& 0 \ \\ 
&\ 0&0& 0&0&0& 0&0&1& 0&0&0&0 \ \\ 
&\ 0&1& 0& 0&0&0& 0& 0& 0& 1& 1&1 \ \\ 
&\ 0& 1& 0&0&0&0&0&0& 1& 0& 1&1 \ \\ 
&\ 0&0&0& 0&0&0&0&0& 0& 0&1&0 \ \\
&\ 0& 0&0& 0&0&0& 0&0&0&0&0& 1 \ \\
\end{block}
\end{blockarray}
\label{Smatrix}
\end{equation}}
All invertible $N\times N$ matrices whose entries take values in $\mathbb{Z}_2$ are valid choices. However, the above choice is the one which best illuminates the geometry of the underlying compactification. Choosing $S=I_N$, the $N$--dimensional identity matrix, would render the translation trivial and the form of $Z\smb{a&k&\cdots}{b&l&\cdots}$ and  $(-1)^{\Psi\smb{a&k&\cdots}{b&l&\cdots}}$ would match that of  $Z\smb{\bm{\alpha}}{\bm{\beta}}$ and $C\smc{\bm{\alpha}}{\bm{\beta}}$, respectively, modulo some subtleties we discuss in the following section. 

Once $S$ is specified, the partition function is written in its index form and we can start making the connection between the GGSO Phases $\CC{\bm{\alpha}}{\bm{\beta}}$ and the modular invariant phase $\Psi$. We assume that $\Psi$ can be expressed as a polynomial in the summation variables. Then, two--loop modular invariance imposed on the GGSO phases via the rule
\begin{equation}\label{TwoLoopMI}
    \CC{\bm{v_i}}{\bm{v_j}+\bm{v_k}} = \delta_{\bm{v_i}} \CC{\bm{v_i}}{\bm{v_j}} \CC{\bm{v_i}}{\bm{v_k}},
\end{equation}
implies that $\Psi\smb{a&k&H_i&h_i&P_i}{b&l&G_i&g_i&Q_i}$ is at most second order in its variables. Moreover, the presence of $\delta_{\bm{v_i}}$ restricts the first--order terms. That is $\Psi$ must include a term $a$ and cannot contain other terms like it. More precisely, (\ref{TwoLoopMI}) implies 
\begin{aline}
     \begin{cases}
         \Psi \ni a,\\
         \Psi \not\owns k,h_i,H_i,P_i,
     \end{cases}
\end{aline}
where we take "$\in$" to mean a term in the sum. These conditions can be implemented in a compact form by requiring the phase to be of the form
\begin{equation}\label{TwoLoopEq}
    \Psi\smb{a&k&H_i&h_i&P_i}{b&l&G_i&g_i&Q_i} = a+\beta_i\Delta_i+\Gamma_i\Omega_{ij}\Delta_j,
\end{equation}
where we defined
\begin{aline}
    \Gamma&=(a,\ k,\ H_1,\ H_2,\ H_3,\ H_4,\ H_5,\ H_6,\ h_1,\ h_2,\ P_1,\ P_2), \\
    \Delta&=(b,\ l,\ G_1,\ G_2,\ G_3,\ G_4,\ G_5,\ G_6,\ g_1,\ g_2,\ Q_1,\ Q_2),
\end{aline}
to be the vectors containing top and bottom indices respectively.

We now impose one--loop modular invariance by requiring that the partition function (\ref{PFTransEq}) remains invariant under $S$ and $T$--transformations, under which the theta functions transform as
\begin{aline}
    S:& \;\;\; \vth\smb{a}{b} \; \longrightarrow \; e^{i \pi ab/2} \; \vth\smb{b}{-a},\\
    T:& \;\;\; \vth\smb{a}{b} \; \longrightarrow \; e^{i \pi a(a-2)/4 } \; \vth\smb{a}{a+b-1}.
\end{aline}
By using a compact notation for the theta and eta function terms as in \eqref{PFTransEq}, i.e. 
\begin{equation}
    Z_F=\frac{1}{2^2}\sum_{\substack{a,k\\b,l}} \; \frac{1}{2^{10}}\sum_{\substack{H_i,h_i,P_i\\G_i,g_i,Q_i}} (-1)^{\Psi\smb{a&k&H_i&h_i&P_i}{b&l&G_i&g_i&Q_i}} Z\smb{a&k&H_i&h_i&P_i}{b&l&G_i&g_i&Q_i},
\end{equation}
we can express the modular transformations more readily. In particular, under modular transformations
\begin{gather*}
   Z\smb{a&k&H_i&h_i&P_i}{b&l&G_i&g_i&Q_i} \overset{S}{\longrightarrow} Z\smb{b&l&G_i&g_i&Q_i}{-a&-k&-H_i&-h_i&-P_i},\\
   Z\smb{a&k&H_i&h_i&P_i}{b&l&G_i&g_i&Q_i} \overset{T}{\longrightarrow} (-1)^{1+a+P_1+P_2} \, Z\smb{a&k&H_i&h_i&P_i}{a+b-1&k+l-1&H_i+G_i-1&h_i+g_i&P_i+Q_i},
\end{gather*}
where the extra factor of $-1$ in the $T$--transformation comes from the $\eta$--functions. By noting that the phase $\Psi$ transforms trivially, as it is just a constant factor, we can conclude that to be modular invariant the phase must satisfy
\begin{aline}\label{STPsi}
     \Psi\smb{a&k&H_i&h_i&P_i}{b&l&G_i&g_i&Q_i} &\overset{S}{=} \Psi\smb{b&l&G_i&g_i&Q_i}{-a&-k&-H_i&-h_i&-P_i},\\
     \Psi\smb{a&k&H_i&h_i&P_i}{b&l&G_i&g_i&Q_i} &\overset{T}{=} 1+a+P_1+P_2+\Psi\smb{a&k&H_i&h_i&P_i}{a+b-1&k+l-1&H_i+G_i-1&h_i+g_i&P_i+Q_i}.
\end{aline}
The first equation, i.e. $S$--invariance, shows that $\Psi$ must be symmetric under the exchange of lower and upper indices, which together with (\ref{TwoLoopEq}), implies that
\begin{equation}\label{OmegaSRules}
    \Psi\smb{a&k&H_i&h_i&P_i}{b&l&G_i&g_i&Q_i} = a+b+\Gamma_i\Omega_{ij}\Delta_j,
\end{equation}
with $\Omega_{ij}=\Omega_{ji}$. Implementing the condition for $T$--invariance in (\ref{STPsi}) further restricts the form of $\Omega$ imposing the conditions on its elements
\begin{aline}\label{OmegaTRules}
    \sum_{\substack{j=1\\j \neq i}}^8 \Omega_{ij} &= 0 \quad &\text{for} \quad i=2,\dots,8,\\
    \sum_{j=1}^8 \Omega_{ij} &= \Omega_{ii}  \quad &\text{for} \quad i=9,10,\\
    \sum_{j=1}^8 \Omega_{ij} &= 1 +\Omega_{ii}  \quad &\text{for} \quad i=11,12,
\end{aline}
where all equalities are understood modulo 2. These fix a further 11 components of $\Omega_{ij}$. Together with the condition from $S$--invariance $\Omega_{ij}=\Omega_{ji}$, we are left with $(12^2/2+12/2)-11=67$ independent choices for the $\Omega_{ij}$. This precisely matches the number of independent GGSO phases for a 12 basis vector model\footnote{We can either count 66 or 67 independent GGSO phases depending on whether we specify that the unimportant phase $\CC{\mathds{1}}{\mathds{1}}$, generating an overall chirality, is fixed or not.}. 

What we achieved here is precisely the derivation of the modular invariance conditions, for the phase $\Psi$. These are analogous to the well--known conditions on the GGSO coefficients in the fermionic formulation. All remaining independent components of $\Omega$ can be freely chosen as $\Omega_{ij}\in \{0,1\}$ with each choice giving a new consistent model. 

\subsection*{The Translation}

We have found a consistent modular invariant way of representing a model in terms of a phase $\Psi$, what remains is to find a translation between the GGSO phases and $\Psi$ as set out in (\ref{PFTransEq}). With the above setup, this means finding a correspondence between the independent GGSO phases  $\CC{\bm{\alpha}}{\bm{\beta}}$ and the matrix elements $\Omega_{ij}$. We have already established that the number of these elements is in agreement on both sides and both quantities perform the same role so such a translation should be possible in principle.

To make the connection, one has to notice that the forms of the theta functions on the left and right--hand sides of (\ref{PFTransEq}) do not match. In particular the expression 
\begin{equation}\label{TransRHS}
    \sum_{\substack{a,k,\cdots\\b,l,\cdots}} (-1)^{\Psi\left[\begin{smallmatrix}a&k&\cdots\\b&l&\cdots\end{smallmatrix}\right]} \;\,Z\left[\begin{matrix}a&k&\cdots\\b&l&\cdots\end{matrix}\right]
\end{equation}
involves theta functions which may take arguments such as $\vth\smb{1}{-1},\vth\smb{3}{0},...$ not permitted on the free fermionic side where the arguments are either 0 or 1. We can, however, use the periodicity properties of the theta functions
\begin{aline}\label{ThetaPeriod2}
   \vth\smb{a+2}{b} =& \,\vth\smb{a}{b}\\
   \vth\smb{a}{b+2} =& \,e^{i \pi a} \vth\smb{a}{b},
\end{aline}
to rewrite \eqref{TransRHS} in terms of the standard theta functions. This will allow for consistent term--by--term matching. By denoting the `fundamental' form of the theta functions as
\begin{equation}\label{ThetaF}
\vth_f\smb{a}{b} \equiv \vth\smb{a \mod 2}{b \mod 2},
\end{equation}
we can find equations using \eqref{ThetaPeriod2} that help bring all theta functions to this reduced form, e.g.
\begin{aline}\label{PeriodRules}
    \vth\smb{a+h_1}{b+g_1} &= (-1)^{(a+h_1)bg_1} \; \vth_f\smb{a+h_1}{b+g_1}\\
    \vth\smb{a-h_1-h_2}{b-g_1-g_2} &= (-1)^{(a-h_1-h_2)(g_1+g_2+bg_1+bg_2+g_1g_2)} \; \vth_f\smb{a-h_1-h_2}{b-g_1-g_2}.
\end{aline}
These relations can always be found by writing $\vth\smb{a&\cdots}{b&\cdots}=(-1)^{F(a,b,\cdots)}\vth_f\smb{a&\cdots}{b&\cdots}$, with $F(a,b,\cdots)$ a suitably general polynomial, and restricting the form of $F$ by requiring \eqref{ThetaPeriod2} to hold. 

Utilising these expressions, we can rewrite the right--hand side of (\ref{PFTransEq}),  fully in terms of the `reduced' theta functions as
\begin{equation}\label{PFTransEqF}
    Z_F=\sum_{\substack{a,k,\cdots\\b,l,\cdots}} (-1)^{\chi\smb{a&k&\cdots}{b&l&\cdots}+\Psi\smb{a&k&\cdots}{b&l&\cdots}} \;\,Z_f\left[\begin{matrix}a&k&\cdots\\b&l&\cdots\end{matrix}\right],
\end{equation}
where we defined the compensating phase factor $\chi$. For our specific model it is given by
\begin{equation}
    \chi\smb{a&k&H_i&h_i&P_i}{b&l&G_i&g_i&Q_i} = (a+k)(g_1+g_2+g_1g_2)+(b+l)(h_1g_2+h_2g_1),
\end{equation}
which enforces the rules \eqref{PeriodRules}. Here, by $Z_f$ we denote that all theta functions have been brought to their mod 2 form as written in (\ref{ThetaF}). This compensating phase is crucial for the matching of the partition functions.

We are now ready to make the connection between the two formalisms.  To compare the two sides of (\ref{PFTransEq}) we must re--express the GGSO matrix $C$ in the form
\begin{equation} \label{CijGij}
    C_{ij}=(-1)^{G_{ij}},
\end{equation}
this allows for a direct comparison of $\Psi$ and $G$. Furthermore, it will be convenient to separate $\Psi$ into its first and second order terms, that is we define
\begin{equation}\label{PsiDef}
    \Psi\smb{a&k&H_i&h_i&P_i}{b&l&G_i&g_i&Q_i} = a+b+\Gamma_i\Omega_{ij}\Delta_j\coloneqq a+b+  \Phi\smb{a&k&H_i&h_i&P_i}{b&l&G_i&g_i&Q_i}.
\end{equation}
We can now express the factor of $a+b+\chi$ in the basis formed by the basis vectors \eqref{SO10basis} as a matrix $P$ whose elements are
\begin{equation}\label{PMat}
    P_{ij} = \{a+b+ \chi\smb{a&k&\cdots}{b&l&\cdots} \,|\; \Gamma_k = S_{ik} \text{ and } \Delta_k = S_{jk} \}.
\end{equation}
All that remains is to express $\Phi$, i.e. $\Omega$, in the same basis so we can equate the two. We can do this by noticing that
\begin{equation}
   \{\Phi\smb{a&k&\cdots}{b&l&\cdots} \,|\; \Gamma_k = S_{ik} \text{ and } \Delta_k = S_{jk} \} = S_{ik}\,\Omega_{kl}\,S_{jl} = S\,\Omega\,S^T,
\end{equation}
and so $\tilde{\Omega}=S\,\Omega\,S^T$ is the phase expressed in the basis formed by the basis vectors of the free fermionic model. Since all quantities are now expressed in the same basis we can write down the equality which implements the translation, namely
\begin{equation}\label{TransSol}
    G+P=S\,\Omega\,S^T,
\end{equation}
where the equality is understood modulo 2. Solving the above equation means finding values for all $\Omega_{ij}$ and so fixing $\Omega$. Once the solution is found to the linear system, the final phase can be expressed using (\ref{PsiDef}), that is 
\begin{equation}\label{PsiSol}
    \Psi\smb{a&k&H_i&h_i&P_i}{b&l&G_i&g_i&Q_i} = a+b+\Gamma\,\Omega\,\Delta. 
\end{equation}
This gives a precise one--to--one correspondence between the modular invariant phase and the GGSO matrix.

It is important to note that the above methods only cover the case for real boundary conditions, i.e. `$\mathbb{Z}_2$--models' such that the fermions are either R or NS. This, in turn, implies that all GGSO phases are real. It is, however, possible to generalise this construction to allow for more general choices of boundary condition vectors and GGSO matrices. 

\section{Scherk--Schwarz and T--duality Conditions} \label{appendix:SS-T-dualityConditions}

In this section we will explicitly show how the techniques developed in 
sections \ref{section:SUSYbreak} and \ref{section:T-duality} can be applied in order to check for SSS and T--duality. In particular, we will use the two models of section \ref{results}, specified by the GGSO phase configurations (\ref{SSmodelGGSO}) and (\ref{GGSOModel2}). \\

The condition for SSS is that for $T_2 \rightarrow \infty$ the partition function vanishes so SUSY is restored, which can happen only if a Jacobi identity is realised. This identity in (\ref{SS-PartitionFunction}) must hold for all indices fixed, except for $H_1,H_2,G_1,G_2$ which do not appear as arguments in the theta functions, since the moduli--dependent lattice $\Gamma^{(1)}_{2,2}\smbar{H_1&H_2}{G_1&G_2}{h_1}{g_1}$ is set to $1$. \\ 
Given the GGSO phases, the phase $\Phi$ in the partition function (\ref{PF}) can be calculated using (\ref{PsiSol}) following the discussion of Appendix \ref{appendix:translation}. For the model (\ref{SSmodelGGSO}) this phase is given by
\begin{align}
\begin{split}
    \Phi \smb{a&k&H_i&h_1&h_2&P_i}{b&l&G_i&g_1&g_2&Q_i} = & \;
    b (a+H_1+h_2+H_2+P_1) \\
    & +l (h_2+k+P_1)\\
    & +G_1 (a+h_1+H_2+H_3+H_4+P_1+P_2)\\
    & +G_2 (a+H_1+h_2+H_2+P_1)\\
    & +G_3 (H_1+H_6)\\
    & +G_4 (H_1+H_4+H_6)\\
    & +G_6 (h_1+h_2+H_3+H_4+H_6)\\
    & +g_1 (H_1+H_6+P_1)\\
    & +g_2 (a+H_2+H_6+k+P_1)\\
    & +Q_1 (a+h_1+H_1+h_2+H_2+k)\\
    & +Q_2 (H_1+P_2) .
\label{eq:phase-SS-Appendix}
\end{split}
\end{align}
Then, as stated in section \ref{section:SUSYbreak}, the Jacobi identity holds only if the following condition is satisfied
\begin{equation} 
\label{SS-condition-Appendix}
\sum_{\substack{H_1, H_2\\G_1, G_2}} (-1)^{\Phi\smb{0&k&H_i&0&0&P_i}{0&l&G_i&0&0&Q_i}} = \sum_{\substack{H_1, H_2\\G_1, G_2}} (-1)^{\Phi\smb{1&k&H_i&0&0&P_i}{0&l&G_i&0&0&Q_i}} = \sum_{\substack{H_1, H_2\\G_1, G_2}} (-1)^{\Phi\smb{0&k&H_i&0&0&P_i}{1&l&G_i&0&0&Q_i}}.
\end{equation}
Actually it is sufficient to prove just the first equality, since due to modular invariance the phase $\Phi \smb{a&k&H_i&h_1&h_2&P_i}{b&l&G_i&g_1&g_2&Q_i}$ is the same by exchanging lower and upper indices. Then, by performing the lattice sum, it is easy to check that the equality holds. 

Alternatively, it can also be check directly by inspecting the phase $\Phi\smb{a&k&H_1&H_2&H_i&0&0&P_i}{b&l&G_1&G_2&G_i&0&0&Q_i}$, where we have distinguished $H_1, H_2$ from the other $H_i$'s to render the discussion clearer. 
For $\Phi\smb{0&k&H_1&H_2&H_i&0&0&P_i}{0&l&G_1&G_2&G_i&0&0&Q_i}$ set the indices $H_1$ and $H_2$ to zero. If there are some combinations of $H_1 ^*$ and $H_2 ^*$ for $\Phi\smb{1&k&H_1 ^*&H_2 ^*&H_i&0&0&P_i}{0&l&G_1&G_2&G_i&0&0&Q_i}$ such that the following equality holds
\begin{equation}
\label{eq:SS-explicit-Phase-Appendix}
    \Phi\smb{0&k&0&0&H_i&0&0&P_i}{0&l&G_1&G_2&G_i&0&0&Q_i} = \Phi\smb{1&k&H_1 ^*&H_2 ^*&H_i&0&0&P_i}{0&l&G_1&G_2&G_i&0&0&Q_i},
\end{equation}
then the condition (\ref{SS-condition-Appendix}) will be satisfied and the Jacobi identity will hold. In particular, for the phase (\ref{eq:phase-SS-Appendix}) it is easy to check that the equality is satisfied setting $H_1 ^* = 0$, $H_2 ^* = 1$. Once this holds, for all other combinations of $H_1, H_2$ the equality will hold according to
\begin{equation}
    \Phi\smb{0&k&0+i&0+j&H_i&0&0&P_i}{0&l&G_1&G_2&G_i&0&0&Q_i} = \Phi\smb{1&k&H_1 ^* +i&H_2 ^* +j&H_i&0&0&P_i}{0&l&G_1&G_2&G_i&0&0&Q_i}.
\end{equation}
This second way of checking for SSS is particularly convenient in terms of time efficiency, being easy to implement in a computational program. \\
We now check whether T--duality is preserved for this model, \textit{i.e.} if condition (\ref{eq:T-duality-condition}) holds. To do this, we check that
\begin{aline}  \sum_{\substack{H_1, H_2\\G_1, G_2}} (-1)^{\Phi\smb{0&k&H_i&0&0&P_i}{0&l&G_i&0&0&Q_i}+H_1G_1 +H_2 G_2} & = \sum_{\substack{H_1, H_2\\G_1, G_2}} (-1)^{\Phi\smb{1&k&H_i&0&0&P_i}{0&l&G_i&0&0&Q_i}+H_1G_1 +H_2 G_2} \\
& = \sum_{\substack{H_1, H_2\\G_1, G_2}} (-1)^{\Phi\smb{0&k&H_i&0&0&P_i}{1&l&G_i&0&0&Q_i}+H_1G_1 +H_2 G_2} 
\end{aline}
is satisfied. As discussed in subsection \ref{section:T-duality}, due to modular invariance, just the first equality is sufficient. Performing the sum, we can see that the T--duality condition does not hold. Also using (\ref{eq:SS-explicit-Phase-Appendix}) modified with the additional T--dual contribution
\begin{equation}
    \Phi\smb{0&k&0&0&H_i&0&0&P_i}{0&l&G_1&G_2&G_i&0&0&Q_i} = \Phi\smb{1&k&H_1 ^*&H_2 ^*&H_i&0&0&P_i}{0&l&G_1&G_2&G_i&0&0&Q_i} + H_1 ^* G_1  + H_2 ^* G_2
\end{equation}
it is easy to check that there is no possible combination of $H_1 ^*, H_2 ^*$ such that the equality is preserved. 

We thus conclude that for this particular model defined by the GGSO phases (\ref{SSmodelGGSO}) the potential will exhibit an SSS breaking with broken T--duality, such that $V(T_2 \rightarrow \infty) \rightarrow 0$ and $V(T_2 \rightarrow 0) \not\to 0$. This behavior of the potential is demonstrated in Figure \ref{figure:SSUpliftedPot}. \\

\noindent
For the second model specified by the GGSO phases (\ref{GGSOModel2}), the phase $\Phi$ is given by
\begin{align}
\begin{split}
    \Phi \smb{a&k&H_i&h1&h2&P_i}{b&l&G_i&g1&g2&Q_i} = & \;
    b (a+h_2+P_2) \\
    & +l (H_2+H_3+H_5+H_6+P_1) \\
    & +G_1 (H_1+H_2+H_3+P_1+P_2) \\
    & +G_2 (H_1+h_2+H_2+H_3+H_4+H_5+H_6+k) \\
    & +G_3 (H_1+H_2+H_4+H_5+H_6+k+P_1+P_2) \\
    & +G_4 (h_1+h_2+H_2+H_3+H_4+P_1+P_2) \\
    & +G_5 (h_1+H_2+H_3+H_6+k+P_1) \\
    & +G_6 (h_2+H_2+H_3+H_5+H_6+k+P_1+P_2) \\
    & +g_1 (h_2+H_4+H_5+P_1) \\
    & +g_2 (a+h_1+H_2+H_4+H_6+P_1) \\
    & +Q_1 (h_1+H_1+h_2+H_3+H_4+H_5+H_6+k+P_2) \\
    & +Q_2 (a+H_1+H_3+H_4+H_6+P_1+P_2).
\end{split}
\end{align}
For this model it can be checked that 
the SSS condition (\ref{SS-condition-Appendix}), or equivalently condition (\ref{eq:SS-explicit-Phase-Appendix}), is not satisfied which implies that the potential diverges at infinity, $V(T_2 \rightarrow \infty) \rightarrow \pm \infty$. The shape of the potential in Figure \ref{figure:UpliftedPot} shows what we expect. 


\section{Chiral Sector Analysis} \label{appendix:ChiralSec}
In addition to the sectors $\bm{F}^{1,2,3}_{pqrs}$ discussed in section \ref{section:chiralsecs}, the following sectors also give rise to massless states transforming under spinorial representations with chirality under the $U(1)_{1,2,3}$ gauge factors
\begin{align}
\begin{split}
    \bm{F}^4_{pqrs}&= \Sv+\bv{1}+\x+\z{1}+p\e{3}+q\e{4}+r\e{5}+s\e{6}\\
    \bm{F}^5_{pqrs}&=\Sv+\bv{2}+\x+\z{1}+p\e{1}+q\e{2}+r\e{5}+s\e{6}\\
    \bm{F}^6_{pqrs}&=\Sv+\bv{3}+\x+\z{1}+p\e{1}+q\e{2}+r\e{3}+s\e{4}\\
    \bm{F}^7_{pqrs}&=\Sv+\bv{1}+\x+\z{2}+p\e{3}+q\e{4}+r\e{5}+s\e{6}\\
    \bm{F}^8_{pqrs}&=\Sv+\bv{2}+\x+\z{2}+p\e{1}+q\e{2}+r\e{5}+s\e{6}\\
    \bm{F}^9_{pqrs}&=\Sv+\bv{3}+\x+\z{2}+p\e{1}+q\e{2}+r\e{3}+s\e{4},
\end{split}
\end{align}
which have the following projecting sets
\begin{align}\label{F456Upsilons}
    \begin{split}
        \Upsilon(\bm{F}^4_{pqrs})&=\{\bm{z_2},\bm{e_1},\bm{e_2}\}\\
        \Upsilon(\bm{F}^5_{pqrs})&=\{\bm{z_2},\bm{e_3},\bm{e_4}\}\\
        \Upsilon(\bm{F}^6_{pqrs})&=\{\bm{z_2},\bm{e_5},\bm{e_6}\}\\
        \Upsilon(\bm{F}^7_{pqrs})&=\{\bm{z_1},\bm{e_1},\bm{e_2}\}\\
        \Upsilon(\bm{F}^8_{pqrs})&=\{\bm{z_1},\bm{e_3},\bm{e_4}\}\\
        \Upsilon(\bm{F}^9_{pqrs})&=\{\bm{z_1},\bm{e_5},\bm{e_6}\},
    \end{split}
\end{align}
and chirality operators 
\begin{align}\label{Chi456}
    \begin{split}
        \chi(\bm{F}^4_{pqrs})&=\text{ch}(\bar{\eta}^{2})+ \text{ch}(\bar{\eta}^{3})= \\ & = - \ \CC{\bm{F}^{4}_{pqrs}}{\Sv+\bv{2}+(1-r)\e{5}+(1-s)\e{6}}^* - \ \CC{\bm{F}^{4}_{pqrs}}{\Sv+\bv{3}+(1-p)\e{3}+(1-q)\e{4}}^* \\
        \chi(\bm{F}^5_{pqrs})&=\text{ch}(\bar{\eta}^{1})+ \text{ch}(\bar{\eta}^{3})= \\ & = - \ \CC{\bm{F}^{5}_{pqrs}}{\Sv+\bv{1}+(1-r)\e{5}+(1-s)\e{6}}^* - \ \CC{\bm{F}^{5}_{pqrs}}{\Sv+\bv{3}+(1-p)\e{1}+(1-q)\e{2}}^* \\
        \chi(\bm{F}^6_{pqrs})&=
        \text{ch}(\bar{\eta}^{1})+ \text{ch}(\bar{\eta}^{2})= \\ & = - \ \CC{\bm{F}^{6}_{pqrs}}{\Sv+\bv{1}+(1-r)\e{3}+(1-s)\e{4}}^* - \ \CC{\bm{F}^{6}_{pqrs}}{\Sv+\bv{2}+(1-p)\e{1}+(1-q)\e{2}}^* \\
        \chi(\bm{F}^7_{pqrs})&=\text{ch}(\bar{\eta}^{2})+ \text{ch}(\bar{\eta}^{3})= \\ & = - \ \CC{\bm{F}^{7}_{pqrs}}{\Sv+\bv{2}+(1-r)\e{5}+(1-s)\e{6}}^* - \ \CC{\bm{F}^{7}_{pqrs}}{\Sv+\bv{3}+(1-p)\e{3}+(1-q)\e{4}}^* \\
        \chi(\bm{F}^8_{pqrs})&=\text{ch}(\bar{\eta}^{1})+ \text{ch}(\bar{\eta}^{3})= \\ & = - \ \CC{\bm{F}^{8}_{pqrs}}{\Sv+\bv{1}+(1-r)\e{5}+(1-s)\e{6}}^* - \ \CC{\bm{F}^{8}_{pqrs}}{\Sv+\bv{3}+(1-p)\e{1}+(1-q)\e{2}}^* \\
        \chi(\bm{F}^9_{pqrs})&=
        \text{ch}(\bar{\eta}^{1})+ \text{ch}(\bar{\eta}^{2})= \\ & = - \ \CC{\bm{F}^{9}_{pqrs}}{\Sv+\bv{1}+(1-r)\e{3}+(1-s)\e{4}}^* - \ \CC{\bm{F}^{9}_{pqrs}}{\Sv+\bv{2}+(1-p)\e{1}+(1-q)\e{2}}^* ,
    \end{split}
\end{align}

Further to these sectors, the following can give rise to massless states when accompanied by one right--moving Neveu--Schwarz oscillator given by
\begin{align}
\begin{split}
\bm{V}^1_{pqrs}&=\Sv+\bv{1}+\x+p\e{3}+q\e{4}+r\e{5}+s\e{6}\\
\bm{V}^2_{pqrs}&=\Sv+\bv{2}+\x+p\e{1}+q\e{2}+r\e{5}+s\e{6}\\
\bm{V}^3_{pqrs}&=\Sv+\bv{3}+\x+p\e{1}+q\e{2}+r\e{3}+s\e{4},
\end{split}
\end{align}
which have the following projecting sets
\begin{align}\label{VUpsilons}
    \begin{split}
        \Upsilon(\bm{V}^1_{pqrs})&=\{\bm{z_1},\bm{z_2},\bm{e_1},\bm{e_2}\}\\
        \Upsilon(\bm{V}^2_{pqrs})&=\{\bm{z_1},\bm{z_2},\bm{e_3},\bm{e_4}\}\\
        \Upsilon(\bm{V}^3_{pqrs})&=\{\bm{z_1},\bm{z_2},\bm{e_5},\bm{e_6}\},
    \end{split}
\end{align}
and chirality operators 
\begin{align}\label{ChiV}
    \begin{split}
        \chi(\bm{V}^1_{pqrs})&=
        \text{ch}(\bar{\eta}^{2})+ \text{ch}(\bar{\eta}^{3})= \\ & = -  \ \CC{\bm{V}^{1}_{pqrs}}{\Sv+\bv{2}+(1-r)\e{5}+(1-s)\e{6}}^* - \ \CC{\bm{V}^{1}_{pqrs}}{\Sv+\bv{3}+(1-p)\e{3}+(1-q)\e{4}}^* \\
        \chi(\bm{V}^2_{pqrs})&=
        \text{ch}(\bar{\eta}^{1})+ \text{ch}(\bar{\eta}^{3})= \\ & = -  \ \CC{\bm{V}^{2}_{pqrs}}{\Sv+\bv{1}+(1-r)\e{5}+(1-s)\e{6}}^* - \ \CC{\bm{V}^{2}_{pqrs}}{\Sv+\bv{3}+(1-p)\e{1}+(1-q)\e{2}}^* \\
        \chi(\bm{V}^3_{pqrs})&=
        \text{ch}(\bar{\eta}^{1})+ \text{ch}(\bar{\eta}^{2})= \\ & = -  \ \CC{\bm{V}^{3}_{pqrs}}{\Sv+\bv{1}+(1-r)\e{3}+(1-s)\e{4}}^* - \ \CC{\bm{V}^{3}_{pqrs}}{\Sv+\bv{2}+(1-p)\e{1}+(1-q)\e{2}}^* .
    \end{split}
\end{align}

\newpage
\printbibliography[heading=bibintoc]
\end{document}